\documentclass[twocolumn,iop]{emulateapj}
\usepackage{apjfonts}
\usepackage{color}

\usepackage{amsmath,graphicx,longtable,hyperref}
\usepackage{natbib,threeparttable,rotating}

\newcommand{\etal}{et al.}
\newcommand{\hbeta}{H{$\beta$}}
\newcommand{\halpha}{H{$\alpha$}}

\newcommand{\CIV}{C{\sevenrm IV}}

\newcommand{\CIII}{C{\sevenrm III]}}

\def\FeII{Fe\,{\sc ii}}
\def\MgII{Mg\,{\sc ii}}
\def\HeII{He\,{\sc ii}}

\newcommand{\OII}{[O{\sevenrm\,II}]}

\newcommand{\NeIII}{[Ne{\sevenrm\,III}]}

\def \OIII {[O\,{\sc iii}]}

\newcommand{\OIIIab}{[O{\sevenrm\,III}]\,$\lambda\lambda$4959,5007}

   \font\sevenrm=cmr7 scaled 1000

\def\kms{{\rm km\,s^{-1}}}

\begin{document}

\title{The Sloan Digital Sky Survey Reverberation Mapping Project: First Broad-line \hbeta\ and \MgII\ Lags at $z\gtrsim 0.3$ from six-Month Spectroscopy}


\author{Yue Shen$^{1,2,3}$, Keith Horne$^4$, C.~J. Grier$^{5,6}$, Bradley M.~Peterson$^{7,8}$, Kelly D.~Denney$^{7,8,9}$, Jonathan R.~Trump$^{5,6,3}$, Mouyuan Sun$^{5,6,10}$, W.~N. Brandt$^{5,6,11}$, Christopher S.~Kochanek$^{7,8}$, Kyle S.~Dawson$^{12}$, Paul J.~Green$^{13}$, Jenny E.~Greene$^{14}$, Patrick B.~Hall$^{15}$, Luis C.~Ho$^{16,17}$, Linhua Jiang$^{16}$, Karen Kinemuchi\altaffilmark{18}, Ian D. McGreer$^{19}$, Patrick Petitjean$^{20}$, Gordon T. Richards$^{21}$, Donald P.~Schneider$^{5,6}$, Michael A.~Strauss$^{14}$, Charling Tao$^{22,23}$, W.~M. Wood-Vasey$^{24}$, Ying Zu$^{25}$, Kaike Pan\altaffilmark{18}, Dmitry Bizyaev\altaffilmark{18}, Jian Ge$^{26}$, Daniel Oravetz\altaffilmark{18},
Audrey Simmons\altaffilmark{18}}

\altaffiltext{1}{Department of Astronomy and National Center for Supercomputing Applications, University of Illinois at Urbana-Champaign, Urbana, IL 61801, USA}
\altaffiltext{2}{Carnegie Observatories, 813 Santa Barbara Street, Pasadena,
CA 91101, USA}
\altaffiltext{3}{Hubble Fellow}
\altaffiltext{4}{SUPA Physics/Astronomy, Univ. of St. Andrews, St. Andrews KY16 9SS, Scotland, UK}
\altaffiltext{5}{Department of Astronomy \& Astrophysics, The Pennsylvania State University, University Park, PA, 16802, USA}
\altaffiltext{6}{Institute for Gravitation and the Cosmos, The Pennsylvania State University, University Park, PA 16802, USA}
\altaffiltext{7}{Department of Astronomy, The Ohio State University, 140 West 18th Avenue, Columbus, OH 43210, USA}
\altaffiltext{8}{Center for Cosmology and AstroParticle Physics, The Ohio State University, 191 West Woodruff Avenue, Columbus, OH 43210, USA}
\altaffiltext{9}{NSF Astronomy \& Astrophysics Postdoctoral Fellow}
\altaffiltext{10}{Department of Astronomy and Institute of Theoretical Physics and Astrophysics, Xiamen University, Xiamen, Fujian 361005, China }
\altaffiltext{11}{Department of Physics, 104 Davey Lab, The Pennsylvania State University, University Park, PA 16802, USA }
\altaffiltext{12}{Department of Physics and Astronomy, University of Utah, Salt Lake City, UT 84112, USA}
\altaffiltext{13}{Harvard-Smithsonian Center for Astrophysics, 60 Garden Street, Cambridge, MA 02138, USA}
\altaffiltext{14}{Department of Astrophysical Sciences, Princeton University, Princeton, NJ 08544, USA}
\altaffiltext{15}{Physics and Astronomy Dept., York University, Toronto, Ontario M3J 1P3, Canada}
\altaffiltext{16}{Kavli Institute for Astronomy and Astrophysics, Peking University, Beijing 100871, China}
\altaffiltext{17}{Department of Astronomy, School of Physics, Peking University, Beijing 100871, China}
\altaffiltext{18}{Apache Point Observatory and New Mexico State University, P.O. Box 59, Sunspot, NM 88349-0059, USA}
\altaffiltext{19}{Steward Observatory, The University of Arizona, 933 North Cherry Avenue, Tucson, AZ 85721-0065, USA}
\altaffiltext{20}{Institut d'Astrophysique de Paris, Universit\'e Paris 6 and CNRS, 98bis Boulevard Arago, 75014 Paris, France}
\altaffiltext{21}{Department of Physics, Drexel University, 3141 Chestnut Street, Philadelphia, PA 19104, USA}
\altaffiltext{22}{Centre de Physique des Particules de Marseille, Aix-Marseille Universit\'e , CNRS/IN2P3, 163, avenue de Luminy - Case 902 - 13288 Marseille Cedex 09, France}
\altaffiltext{23}{Tsinghua Center for Astrophysics, Tsinghua University, Beijing 100084, China}
\altaffiltext{24}{PITT PACC, Department of Physics and Astronomy, University of Pittsburgh, 3941 O'Hara Street, Pittsburgh, PA 15260, USA}
\altaffiltext{25}{McWilliams Center for Cosmology, Department of Physics, Carnegie Mellon University, 5000 Forbes Avenue, Pittsburgh, PA 15213, USA}
\altaffiltext{26}{Astronomy Department, University of Florida, 211 Bryant Space Science Center, Gainesville, FL 32611, USA}

\shorttitle{SDSS-RM: First Lags}


\shortauthors{SHEN ET~AL.}

\begin{abstract}

Reverberation mapping (RM) measurements of broad-line region (BLR) lags in $z>0.3$ quasars are important for directly measuring black hole masses in these distant objects, but so far there have been limited attempts and success given the practical difficulties of RM in this regime. Here we report preliminary results of 15 BLR lag measurements from the Sloan Digital Sky Survey Reverberation Mapping (SDSS-RM) project, a dedicated RM program with multi-object spectroscopy designed for RM over a wide redshift range. The lags are based on the 2014 spectroscopic light curves alone (32 epochs over 6 months) and focus on the \hbeta\ and \MgII\ broad lines in the 100 lowest-redshift ($z<0.8$) quasars included in SDSS-RM; they represent a small subset of the lags that SDSS-RM (including 849 quasars to $z\sim 4.5$) is expected to deliver. The reported preliminary lag measurements are for intermediate-luminosity quasars at $0.3\lesssim z<0.8$, including 9 \hbeta\ lags and 6 \MgII\ lags, for the first time extending RM results to this redshift-luminosity regime and providing direct quasar black hole mass estimates over $\sim$ half of cosmic time. The \MgII\ lags also increase the number of known \MgII\ lags by several-fold, and start to explore the utility of \MgII\ for RM at high redshift. The location of these new lags at higher redshifts on the observed BLR size-luminosity relationship is statistically consistent with previous \hbeta\ results at $z<0.3$. However, an independent constraint on the relationship slope at $z>0.3$ is not yet possible due to the limitations in our current sample. Our results demonstrate the general feasibility and potential of multi-object RM for $z> 0.3$ quasars.

\end{abstract}


\section{Introduction}\label{sec:intro}

Measuring the masses of black holes in quasars and active galactic nuclei (AGN)\footnote{We use the term ``AGN'' to refer to the low-luminosity counterparts of quasars. } is of critical importance to many fundamental problems in AGN physics and phenomenology, the growth of supermassive black holes (SMBHs), and the co-evolution of galaxies and SMBHs. As the primary method to measure active SMBH masses, the reverberation mapping (RM) technique estimates the size of the BLR by measuring the time lag between continuum variations and line responses \citep[][]{Blandford_McKee_1982,Peterson_1993,Peterson_2013}. This technique has been widely practiced in the past two decades, resulting in BLR RM measurements for $\sim 60$ low-redshift ($z<0.3$) AGN and quasars \citep[e.g.,][]{Peterson_etal_1998a,Kaspi_etal_2000,Kaspi_etal_2005,Peterson_etal_2002,
Peterson_etal_2004,Bentz_etal_2009,Bentz_etal_2010,Bentz_etal_2013,Denney_etal_2009a,Denney_etal_2010,Rafter_etal_2011,Rafter_etal_2013, Barth_etal_2011a,Barth_etal_2011b,Barth_etal_2013,Barth_etal_2015,Grier_etal_2012,Du_etal_2014,Du_etal_2015,Hu_etal_2015}, mostly focused on the \hbeta\ line. Beyond $z>0.3$, however, RM results are scarce, given the stringent observational requirements for detecting lags in these distant and faint objects: there have been only a handful of attempts for the most luminous quasars \citep[e.g.,][]{Kaspi_etal_2007,Trevese_etal_2007,Trevese_etal_2014} and only one or two tentative \CIV/\CIII\ lag detections have been reported.

In addition to sample size and redshift coverage, another limitation of the current RM sample is the deficit of lag detections for the \MgII\ line, despite the fact that \MgII\ is one of the most important  lines of RM interest that can be observed in the optical at $0.3<z<2$. Locally, there is only one reliable detection of a \MgII\ lag in NGC 4151 \citep{Metzroth_etal_2006}, and two marginal \MgII\ lag detections in NGC 5548 \citep{Clavel_etal_1991} and NGC 3783 \citep{Reichert_etal_1994}; all were based on UV spectroscopy from space. The prospects of \MgII\ RM in the optical for high-redshift quasars therefore demand a systematic examination. 

To expand the redshift range and AGN/quasar parameter space for RM measurements, and to evaluate the potential of RM on all prominent broad lines, we are conducting a dedicated multi-object RM program, SDSS-RM. The first-year observations were completed in 2014 as an ancillary program of the SDSS-III surveys \citep{Eisenstein_etal_2011}. SDSS-RM spectroscopically monitors a flux-limited ($i_{\rm psf}<21.7$) sample of 849 quasars in a single $7\ {\rm deg^2}$ field with the SDSS-III Baryon Oscillation Spectroscopic Survey (BOSS) spectrograph \citep{Dawson_etal_2013,Smee_etal_2012} on the 2.5~m SDSS telescope \citep{Gunn_etal_2006}, accompanied by dedicated photometric monitoring from a number of ground-based wide-field imagers. With its multiplex advantage, SDSS-RM offers considerably higher efficiency than traditional RM programs executed in a serial mode, and is designed to perform RM for a homogeneous selection of quasars over a wide redshift and luminosity range. The details of the program are described in \citet{Shen_etal_2015a}.

In this work we report initial broad-line lag detections from the first-year SDSS-RM observations. These detections are based on the 6-month spectroscopic data alone and demonstrate the feasibility of lag detections using only carefully-calibrated multi-fiber spectroscopy. The reported detections are for \hbeta\ and \MgII\ only, and serve as a proof-of-concept study of the general feasibility of multi-object RM programs; they do not represent the complete set of lag detections from SDSS-RM, nor do we quantify the completeness of lag detections and selection biases inherent to our program in this work. We also do not consider the important aspects of multiple line detections in the same object. More comprehensive analyses of the SDSS-RM data with different focuses are deferred to future work. Nevertheless, the lag detections reported here represent a significant advance in RM: 1) they form the largest sample of lag detections for $z\gtrsim 0.3$ quasars, whose luminosities overlap with those in the $z<0.3$ RM sample but are much lower than those in the handful of $z>0.3$ quasars previously monitored for RM, thus greatly expanding the luminosity-redshift range for which RM observations have been successful; 2) they comprise the largest sample of \MgII\ lag detections to date, starting to explore the utility of this line for RM purposes. 

We describe the data and technical details of the time series analysis in \S\ref{sec:data}, and present our results in \S\ref{sec:disc}. We conclude in \S\ref{sec:con} with a brief outline for future work. A flat $\Lambda$CDM cosmology with $\Omega_\Lambda=0.7$ ($\Omega_0=0.3$) and $h_0=0.7$ is adopted throughout. By default the reported lags are in the quasar rest-frame. For the ease of discussion, we use ``low-$z$'' to refer to $z<0.3$.

\section{Data and Time Series Analysis}\label{sec:data}

The spectroscopic data were taken during seven dark/grey runs from January to July 2014, and consist of a total of 32 epochs with an average cadence of $\sim 4$ days; each epoch had a typical exposure time of 2 hrs. The spectroscopic data were pipeline-processed as part of the SDSS-III Data Release 12 \citep{dr12}, followed by a custom flux calibration scheme and improved sky subtraction as described in \citet{Shen_etal_2015a}. The improved spectrophotometry has a nominal accuracy of $\sim 5\%$. The wavelength coverage of BOSS spectroscopy is $\sim 3650-10,400~\textrm{\AA}$, with a spectral resolution of $R\sim 2000$. The typical S/N per 69 ${\rm km\,s^{-1}}$ pixel averaged over the $g$ band in a 2-hr exposure is $\sim 4.5$ at $g_{\rm psf}=21.2$.

Next, we perform a spectral refining procedure on the custom flux-calibrated multi-epoch spectra, called ``PrepSpec'', to reduce further the scatter in the flux calibration. As briefly described in section 3.4 of \citet[][]{Shen_etal_2015a}, PrepSpec rescales the flux levels of each individual epoch by optimizing model fits (in parameterized functional forms) to describe the continuum and broad-line variability patterns as functions of time and wavelength, using the fluxes of the narrow emission lines (in particular \OIIIab) as an internal calibrator \citep[e.g.,][]{vw_92}, which are assumed to remain constant over the relatively short monitoring period \citep[e.g.,][]{Peterson_etal_2013}. This procedure improves the calibration of the relative spectrophotometry to $\lesssim 2\%$ for low-redshift quasars with strong narrow emission lines. Given the typical $\sim 10\%$ variability amplitude of the continuum and responding broad line fluxes \citep[e.g.,][]{Peterson_2013}, this additional improvement of the spectrophotometry is essential for detecting BLR lags based on spectroscopy alone. As a byproduct of PrepSpec, we obtained model light curves (LCs) for the desired broad emission line and continuum fluxes, as well as model mean and RMS profiles of the broad lines. We use the rest-frame 5100\,\AA\ continuum flux as the fiducial continuum LC used in the time series analysis, given the proximity of this continuum to the narrow \OIIIab\ lines that are used to calibrate the spectrophotometry. Since the fitting was performed over a large number of pixels, the measurement uncertainties in the model fluxes are typically much smaller than the uncertainties of fluxes directly measured over a narrow range of pixels, as reported in most previous RM work. These PrepSpec outputs form the basis for our following analysis. 

We now describe PrepSpec in further detail. 


\subsection{PrepSpec Analysis}\label{sec:prepspec}

\begin{figure*}
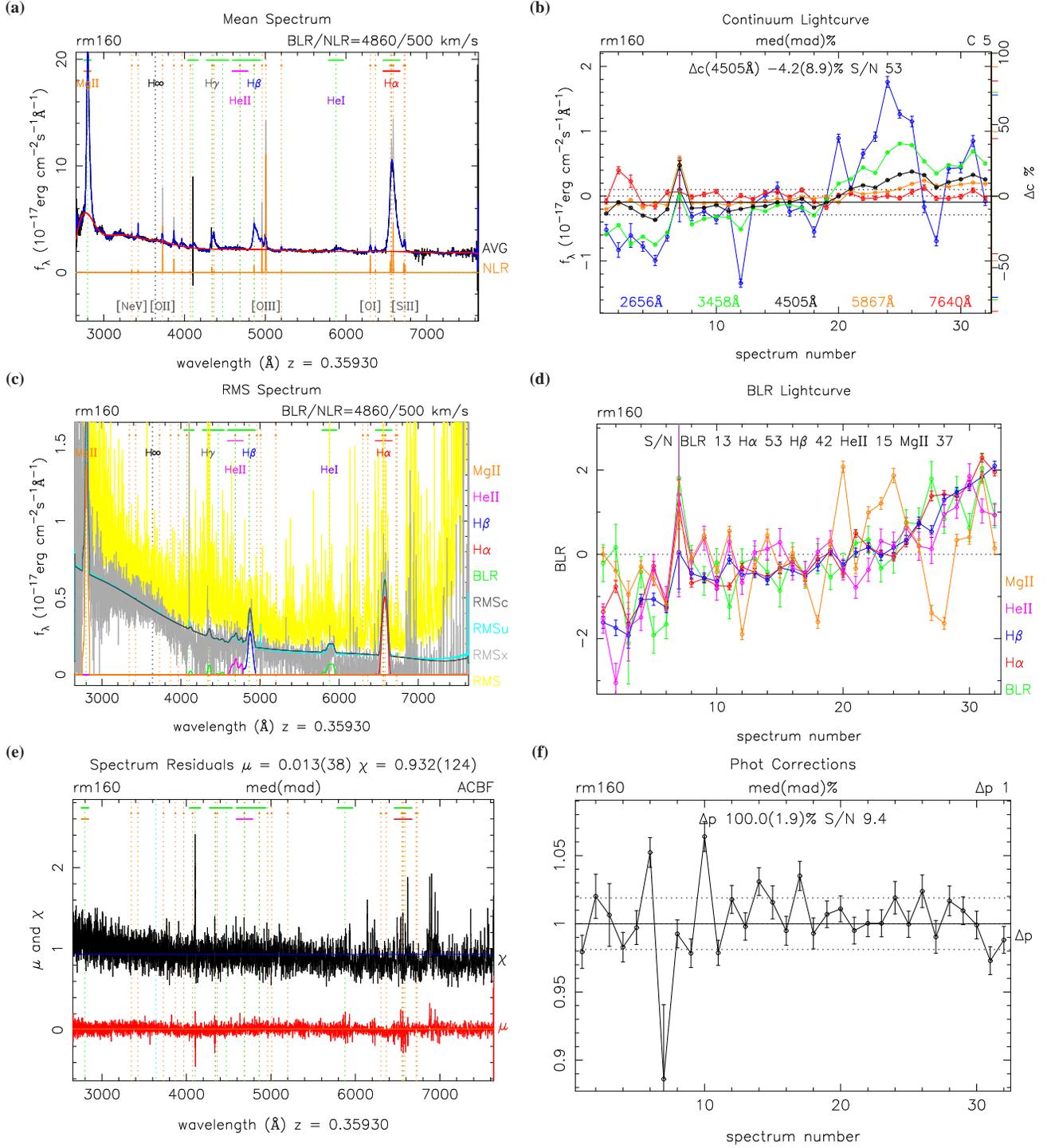

\begin{tabular}{cc}
{\bf (a)}
\includegraphics[angle=270,width=8cm]{rm160_avg_w.eps}
&
{\bf (b)}
\includegraphics[angle=270,width=8cm]{fig1b.eps}
\\
{\bf (c)}
\includegraphics[angle=270,width=8cm]{rm160_blr_w.eps}
&
{\bf (d)}
\includegraphics[angle=270,width=8cm]{rm160_blr_t.eps}
\\
{\bf (e)}
\includegraphics[angle=270,width=8cm]{rm160_ACBF_chi-bias.eps}
&
{\bf (f)}
\includegraphics[angle=270,width=8cm]{fig1f.eps}
\end{tabular}
\caption[] {\small
Results of the PrepSpec analysis of RMID-160.
{\bf (a)} The mean spectrum $A(\lambda)$ decomposed
into $\bar{F}(\lambda)$ and $N(\lambda)$. The grey line is the mean spectrum; the blue line is a spline fit used to isolate the NLR spectrum (sigma-clipped spline fit with narrow line windows masked); the black line is the mean spectrum after subtracting the NLR spectrum; the orange line is the NLR spectrum. 
The narrow-line and broad-line windows are marked
at the top, with vertical dashed lines at the rest wavelengths.
{\bf (b)} Continuum variations $C(\lambda,t)$
evaluated at 5 wavelengths across the spectrum.
{\bf (c)} The RMS spectrum of the raw data (yellow),
and RMSx (grey), the maximum likelihood estimate for the excess RMS (i.e., removing the contribution from the assumed Gaussian measurement errors) after the $p(t)$ corrections.
RMSu (cyan) is the uncalibrated RMS of the model, before the photometric $p(t)$ corrections.
RMSc (dark grey) is the calibrated RMS of the model, after the $p(t)$ corrections. 
{\bf (d)} The broad-line LCs, $L_\ell(t)$,
normalized to mean 0 and RMS 1, and estimated
S/N for detection of variations in each line (``BLR'' collectively refers to all broad lines other than the ones labelled).
{\bf (e)} The mean (red) and RMS (black) of normalized residuals
at each wavelength.
{\bf (f)} The photometric corrections $p(t)$ calibrating time-dependent flux calibration errors.
}
\label{fig:ps}
\end{figure*}

\begin{figure*}
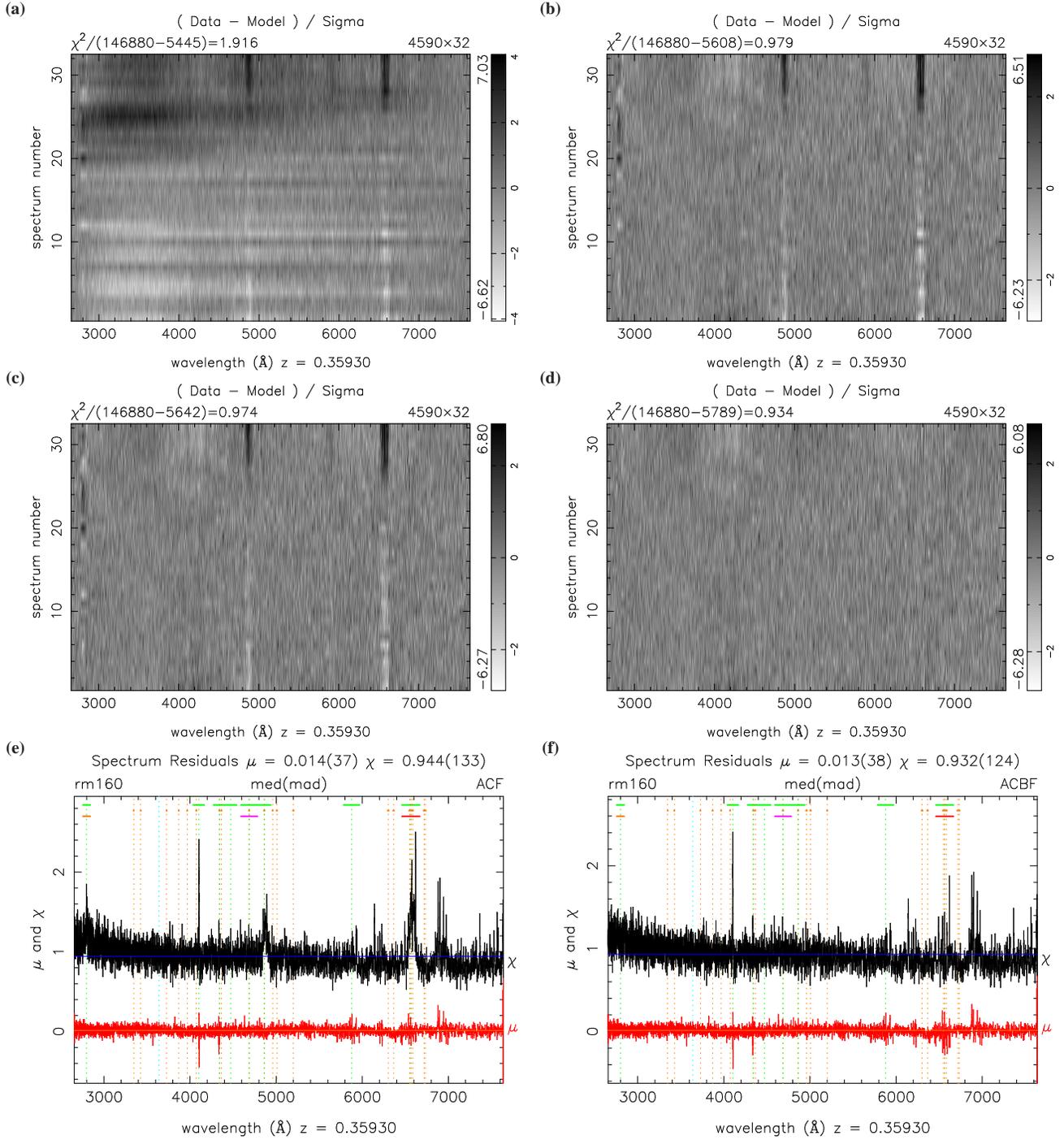

\begin{tabular}{cc}
{\bf (a)}
\includegraphics[angle=270,width=8cm]{rm160_A_eta.eps}
&
{\bf (b)}
\includegraphics[angle=270,width=8cm]{rm160_AC_eta.eps}
\\
{\bf (c)}
\includegraphics[angle=270,width=8cm]{rm160_ACF_eta.eps}
&
{\bf (d)}
\includegraphics[angle=270,width=8cm]{rm160_ACBF_eta.eps}
\\
{\bf (e)}
\includegraphics[angle=270,width=8cm]{rm160_ACF_chi-bias.eps}
&
{\bf (f)}
\includegraphics[angle=270,width=8cm]{rm160_ACBF_chi-bias.eps}
\end{tabular}
\caption[] {\small
Results of PrepSpec analysis of {RMID-160}.
Greyscale maps of normalized residuals $\chi(\lambda,t)$
are shown after fitting {\bf (a)} the mean spectrum $A(\lambda)$,
{\bf (b)} the continuum variations $C(\lambda,t)$ in the final PrepSpec output,
{\bf (c)} the photometric corrections $p(t)$,
and {\bf (d)} the broad-line variations $B(\lambda,t)$.
The mean and RMS of $\chi(\lambda,t)$ over time,
$\mu(\lambda)$ and RMS $\chi(\lambda)$ respectively,
are shown {\bf (e)} before and {\bf (f)} after
fitting the broad-line variations.
These results exhibit clear evidence for intrinsic variations 
in the continuum and in the broad H$\alpha$ and H$\beta$ lines.
}
\label{fig:chimap}
\end{figure*}

PrepSpec aims to improve the calibration of time-resolved spectral data
by fitting a model that includes intrinsic variations in the continuum and
broad emission lines, and corrections for residual calibration errors.
For each SDSS-RM spectrum in any given epoch, our PrepSpec model is
\begin{equation}\label{eqn:model}
\mu(\lambda,t) = p(t) \left[ A(\lambda)
	+ B(\lambda,t)
	+ C(\lambda,t)
	\right]
\ .
\end{equation}
Here $p(t)$ are time-dependent photometric corrections,
$A(\lambda)$ is the average spectrum, and
$B(\lambda,t)$ and $C(\lambda,t)$ are
variations in the BLR spectrum and continuum respectively.
Corrections for residual wavelength shifts and 
spectral blurring were investigated and found not to be needed.
Figs.~\ref{fig:ps} and \ref{fig:chimap} 
illustrate some results of the PrepSpec analysis using one of our objects which showed significant continuum and line variability over the monitoring period (RMID-160).

The average spectrum is decomposed as
\begin{equation}
A(\lambda) = \bar{F}(\lambda) + N(\lambda)
\ .
\end{equation}
Here $\bar{F}(\lambda)$ includes both the continuum and broad-line components. The narrow-line component, $N(\lambda)$,
is isolated using a piecewise cubic spline fit to $A(\lambda)$ as 
a high-pass filter and then multiplying the result by
a window function that is 0 outside and 1 inside a
defined velocity range around a specified list
of narrow emission lines. The narrow-line list includes \OIIIab,
but also many other weaker narrow lines across the spectrum (e.g., Balmer lines, \HeII\ $\lambda$4687, \OIII\ $\lambda$4364, \NeIII\ $\lambda\lambda$3968,3869, \OII\ $\lambda$3728, etc.),
as shown for RMID-160 in Fig.~\ref{fig:ps}a.

The model assumes that the narrow lines have no intrinsic 
variations, so that apparent variations in the data
are interpreted as flux calibration scatter $p(t)$, assumed to be wavelength-independent.
PrepSpec fits $\ln(p(t))$, so that $p(t)$ remains positive, and
normalizes $p(t)$ to a median of 1.
Fig.~\ref{fig:ps}f shows $p(t)$ varying by about 2\%
(median absolute deviation, MAD) for RMID-160, reflecting the precision of our custom flux calibration \citep[on average 5\% in RMS;][]{Shen_etal_2015a}.

Continuum variations are modeled by a polynomial in $\log{\lambda}$ 
with $N_C=5$ time-dependent coefficients:
\begin{equation}
 C(\lambda,t) = \sum_{k=0}^{N_C-1} C_k(t)\,
	\left[ \eta(\lambda) \right]^k
\ ,
\end{equation}
where
$ \eta(\lambda) \equiv
 2\, \log{ \left( \lambda / \sqrt{ \lambda_1\, \lambda_2 }
 \right)} / \log{ \left( \lambda_2/\lambda_1 \right)}
$
varies from $-1$ to $+1$ across the spectrum from wavelengths
$\lambda_1$ to $\lambda_2$.
Fig.~\ref{fig:ps}b shows for RMID-160 the
continuum LCs at 5 wavelengths.
The intrinsic continuum variations are well detected,
generally increasing from red to blue.
Some large-amplitude features on the blue end of the
spectrum are due to residual instrumental 
and flux calibration errors.

The BLR variations are represented by a separable function
for each line:
\begin{equation}\label{eqn:bb}
	B(\lambda,t) = \sum_{\ell=1}^{N_\ell}
	B_\ell(\lambda)\, L_\ell(t)
\ .
\end{equation}
Here $L_\ell(t)$, the LC of line $\ell$, is normalized
to a mean of 0 and RMS of 1, so that
$B_\ell(\lambda)$ is the RMS spectrum of line $\ell$. Optimal scaling of $L_\ell(t)$ provides
pixel-by-pixel estimates of $B_\ell(\lambda)$ with error bars, which are then smoothed with a spline function. 
The assumption that the broad-line flux variations are 
separable in wavelength and time as in Eqn.\ (\ref{eqn:bb}) is of course a simplification that greatly speeds up the 
calculations. It assumes that the variable broad-line emission has a constant velocity profile during the monitoring period, which is not necessarily 
true given that different parts of the BLR reverberate at different times. However, we found this simplified model can fit the variable BLR emission well on an epoch-by-epoch basis, suggesting that more sophisticated models are unnecessary given the level of statistical errors on the flux measurements.  

The RMS spectra and BLR LCs for RMID-160 are
shown in Fig.~\ref{fig:ps}c and d, respectively.
Variations in H$\alpha$ and H$\beta$ are well detected. The \MgII\ line, on the blue edge of the RM-160 spectrum,
exhibits some large-amplitude LC features,
correlating with those in the blue continuum,
that are likely due to residual wavelength-dependent
flux calibration errors that can be diagnosed but
are not calibrated out in the current PrepSpec analysis.

PrepSpec optimizes the model parameters to minimize the $\chi^2$ for the model fit to the data.
The parameters include the mean spectrum
$A(\lambda)$, the BLR profiles $B_\ell(\lambda)$
and the LCs $p(t)$, $C_k(t)$, and $L_\ell(t)$.
The fit begins with initial estimates for $A(\lambda)$ and 
$B_\ell(\lambda)$, and $p(t)=1$.
A series of linear regression fits, one at each time,
adjusts $p(t)$, $C_k(t)$ and $L_\ell(t)$, providing error bars and mutual covariances as well.
Constraints are imposed, normalizing $p(t)$ to 
median 1, and $L_\ell(t)$ to mean 0 and RMS 1.
The spectra $A(\lambda)$ and $B_\ell(\lambda)$ are then
adjusted by optimally scaling the appropriate LCs
to fit the residuals at each wavelength. This also
provides error bar spectra.
Spline fits and windowing constraints 
then serve to decompose $A(\lambda)$, to
smooth $B_\ell(\lambda)$ and to impose the 
appropriate BLR and NLR line windows.
Several line width measures (FWHM, RMS, MAD) are then 
determined for each of the narrow and broad lines,
and the narrow-line and broad-line window widths are
adjusted accordingly.
The above steps are iterated until convergence.

Maps of the residuals normalized by the error per pixel $\chi(\lambda,t)$
and the corresponding $\chi(\lambda)$ spectrum,
as shown for RMID-160 in Fig~\ref{fig:chimap},
provide helpful diagnostics of the fit 
at four stages during optimization of the PrepSpec model,
and exhibit the evidence, or lack thereof, for
intrinsic variations in the lines and continuum.
In Fig~\ref{fig:chimap}a, 
after fitting the mean spectrum $A(\lambda)$, 
the $\chi(\lambda,t)$ map shows both horizontal
stripes due to apparent continuum variations and vertical
stripes due to apparent line variations.
In Fig~\ref{fig:chimap}b, the continuum variations $C(\lambda,t)$
are well modeled and removed from the residuals, leaving 
the line variations.
At this stage the residual flux calibration errors are not yet corrected, and so
apparent variations in the narrow lines are often evident,
typically at a 5\% level \citep[i.e., the spectrophotometry achieved for SDSS-RM;][]{Shen_etal_2015a}.
These are calibrated by optimizing $p(t)$, 
with the resulting $\chi(\lambda,t)$ in Fig~\ref{fig:chimap}c,
and the corresponding $\chi(\lambda)$ spectrum in Fig~\ref{fig:chimap}e,
then exhibiting the evidence for broad-line variations,
in this case evident for H$\alpha$, H$\beta$ and \MgII.
For the final model (i.e., Eqn.\ \ref{eqn:model}), the residuals in Fig~\ref{fig:chimap}d
bear a satisfying resemblance to white noise.
There is excess variance in spectral regions where telluric
sky lines have been subtracted and at the join between the red
and blue parts of the spectrum.

The estimated errors for the continuum and broad-line LCs properly take into account the covariance between the LCs and the photometric correction $p(t)$. The RMS profile of the variable broad-line component is taken as the PrepSpec model $B_\ell(\lambda)$; this approach differs from most earlier RM work, in which an RMS spectrum is generated using the full spectra and then a continuum is subtracted and the line widths are measured from the resulting spectrum (which differs from the true broad-line RMS variations). Most earlier RM work used this traditional approach due to practical difficulties in isolating the continuum and line components in individual epochs, but the most recent RM work has started to construct the line LCs and RMS spectrum by decomposing various components in individual epochs \citep[e.g.,][]{Bian_etal_2010,Park_etal_2012a,Barth_etal_2015,Hu_etal_2015}. What PrepSpec does is similar to the decomposition approach used in the latest studies \citep[e.g.,][]{Barth_etal_2015}, and it is more appropriate and robust than the traditional approach for estimating unbiased broad-line RMS variations, in particular for low quality data. However, the current PrepSpec version does not model the UV and optical \FeII\ complex and host galaxy due to insufficient data quality for most objects in individual epochs. In addition, the assumption that the narrow-line flux remains constant during the monitoring of our program may be violated if the NLR size is comparable to or larger than the 2\arcsec\ fiber diameter, in which case seeing variations (coupled with guiding errors) will induce noticeable changes in the enclosed narrow-line flux. These complications may impact the lag detection in some objects by reducing the correlation signal, and will be investigated systematically in future work with photometric LCs incorporated. 

Given the nature of the SDSS-RM program and target properties, our cadence, spectral S/N, and the quality of the resulting LCs are typically much worse than those in the most recent (traditional single-object) RM work. These circumstances necessitate the usage of PrepSpec to provide robust estimates of the spectral quantities (e.g., fluxes, RMS line profiles, etc.) and their associated uncertainties, as opposed to direct measurements from the spectrum as done in most traditional RM work. 


\subsection{Lag Measurements}\label{sec:lag_det}

We have performed PrepSpec analysis on the 100 lowest-redshift quasars in our sample ($0.116<z<0.782$). In most of these objects we detect significant continuum and broad-line variations. Inspection of the cross-correlation functions (CCF) between continuum and broad-line LCs suggests that a significant fraction ($\sim 30-40\%$) of the 100 low-$z$ targets show evidence of a time lag in at least one of the broad lines. However, the quality of the spectroscopic LCs, coupled with possible correlated errors between the continuum and line LCs measured from spectroscopy, prevent a robust lag detection in many cases. We expect the situation will improve significantly when we incorporate the denser photometric LCs from our accompanying imaging in future work. 

We have noticed that some of the fractional errors on the continuum LCs from PrepSpec are much smaller than $1\%$ in the brightest targets (e.g., photon noise is almost negligible). As we are using the fluxes of the narrow lines to internally calibrate the spectrophotometry (see \S\ref{sec:prepspec}), the ultimate limitation on the precision of the spectroscopic continuum flux measurements lies in the assumptions that narrow line fluxes within the SDSS fibers are constant over the monitoring period and that host starlight contamination is not strongly seeing-dependent. While it may be the case that both the host starlight and the narrow line region are compact enough such that aperture/seeing effects do not introduce additional systematics, it is unlikely that the spectroscopic continuum fluxes can be measured to much better than $1\%$. To test this, we measure the fractional differences in continuum flux measurements for pairs of epochs separated by less than 2 days. There are typically 5-6 such pairs per light curve. The mean fractional differences range from $1-5\%$ with a mean of $2.8\pm 1.0\%$. This sets an upper limit on the true fractional errors of the spectroscopic flux measurements, as AGNs do appear to vary significantly at (or even above) this level on such short timescales \citep[e.g.,][]{Barth_etal_2015}. In fact, this test suggests that PrepSpec in general improved the flux calibration to better than $\sim 2\%$.\footnote{We also compared the spectroscopic continuum LCs with preliminary photometric LCs (only half of the photometric data were processed), and generally found good agreement between the two. This shows promise for our next steps in improving lag measurements with the full SDSS-RM data set (spectroscopy and photometry). } Nevertheless, to be conservative with the LC errors and the lag detections, we inflate the PrepSpec errors (in both continuum and line fluxes) to 3\% of the median flux and use them in our following CCF analysis instead of the nominal PrepSpec errors (unless the latter are already greater than 3\%). We note that in most objects, this detail does not matter at all, as the dominant uncertainty in the lag comes from the sampling of the LC rather than from the LC errors (which are too small to affect the lag measurement). We do notice that using inflated errors leads to larger measurement errors in the lag for some objects, where the errors in both the LC and the lag are likely overestimated.  

For this initial lag study, we focus on the \hbeta\ and \MgII\ lines, which are of the most value in the redshift range considered here. There are 155 LC pairs out of the 100 objects. All time series analyses are based on the spectroscopic LCs output by PrepSpec. To ensure the reported detections are robust, we perform a series of rigorous tests combined with visual inspections, as detailed below. 

First, we require that significant variability is detected in both the continuum and line fluxes over the time span of the LCs. We quantify the LC variability by the amplitudes of the variable part of the continuum and broad-line LCs output by the PrepSpec modeling, normalized by the measurement error ($\rm{S/N_{\rm LC}}$); we only consider LCs with $\rm{S/N_{\rm LC}}>10$ in this study to reduce false positives, but we note that lag detections are possible with a lower threshold in the LC variability \citep[e.g.,][]{Shen_etal_2015a}. We then perform the standard interpolated CCF analysis upon the LCs \citep[e.g.,][]{Gaskell_Peterson_1987} with a grid size of 2 days and identify peaks with a statistical significance $p>0.999$ \citep[e.g.,][sec 2.6]{Bevington_1969,Shen_etal_2015a}. This peak significance quantifies the probability that the detected correlation is not from random fluctuations in uncorrelated time series, and it is particularly useful for quickly removing false positives from large samples of RM data with moderate-quality light curves (such as in SDSS-RM). {We intentionally used the continuum LC for interpolation in the CCF calculation, because the line LC is usually noisier than the continuum LC. But we verified that similar CCF results are obtained by interpolating both the continuum and line LCs and then using the average. }

We measure the lag and its uncertainty following the standard FR/RSS (flux redistribution/random subset selection) Monte Carlo method \citep{Peterson_etal_1998}. For each LC set, we generate 1000 bootstrap trials with added random noise from the adopted flux errors. For each FR/RSS trial we calculate the centroid of the CCF by using all points with $r>0.8r_{\rm max}$ around the peak, where $r$ is the correlation coefficient. We derive the CCF centroid distribution (CCCD) from these FR/RSS trials and use the median of the CCCD as the measured lag $\tau_{\rm cent}$, and the 16th and 84th percentiles of the CCCD as the 1$\sigma$ uncertainty. We only consider a measured lag a detection if it is inconsistent with zero within the 1$\sigma$ uncertainty.

{Using the median of the CCCD as the reported lag differs from using the centroid of the CCF computed from the full LC data. Both approaches are used in the literature, and in the case of sparsely-sampled LCs, the two approaches may produce noticeable differences in the ``best value'' of the measured lag (e.g., RMID-645). However, the formal uncertainties estimated from FR/RSS are generally substantial in such cases. We adopted the median of the CCCD as the reported lag, as this approach is less susceptible to shot noise in sparsely-sampled LC data and provides more symmetric uncertainties around the best value compared to using the CCF centroid derived from the full LC data. }


\begin{figure}
\centering
    \includegraphics[width=0.48\textwidth]{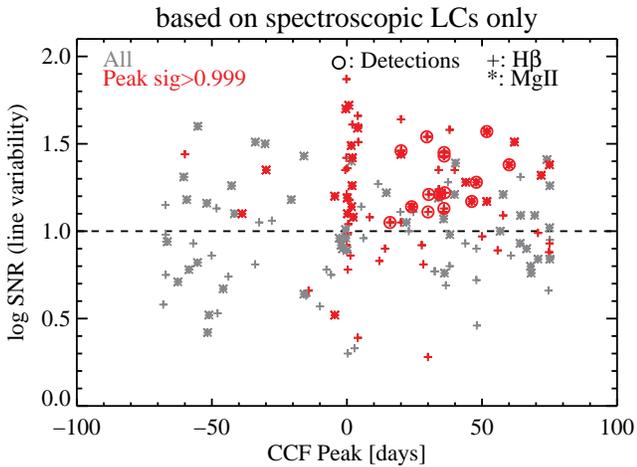}
     \caption{{Diagnosis of lag detections with spectroscopic-only LCs. Gray points show the CCF peaks against the signal-to-noise ratio (SNR) of the line variability for the 100 lowest-redshift quasars in SDSS-RM for which we have processed with PrepSpec, focusing only on the \hbeta\ and \MgII\ lines. Red points show only those peaks with a statistical significance greater than 0.999, thus removing spurious peaks from low-quality LC data. There is an obvious preference towards positive peaks (i.e., lags) in the high-significance peaks, indicating these spectroscopic LCs are meaningful in detecting lags. Objects with larger variability amplitudes in the lines allow more straightforward lag detection (i.e., comparing the gray and red points). There is an excess of zero-lag peaks, which reflects correlated errors in the continuum and line LCs from spectroscopy alone, and/or the difficulty to detect lags shorter than the spectroscopic cadence (a few days). Finally, the red circled points show our reported detections, which have a CCF peak statistical significance greater than 0.999, a SNR in the line variability greater than 10, and a measured lag inconsistent with zero at $>1\sigma$ (see \S\ref{sec:lag_det} for details).}}
    \label{fig:diag}
\end{figure}

We examined the LCs and CCFs for all cases that satisfy the above criteria, and chose the 15 best cases (as judged by the smoothness of the CCF upon visual inspection) to include in this work. While the selection of these detections is by no means complete or objective, it does not affect our basic conclusions. For example, no prior on the expected lag (i.e., from the $R-L$ relation derived from the low-$z$ RM AGN sample) was imposed when we select these detections. As sanity checks, we measure the lags on these LCs using the discrete-correlation-function \citep{Edelson_Krolik_1988} and JAVELIN \citep{Zu_etal_2011}, and found consistent lag measurements.\footnote{We found that in a few cases, JAVELIN reports substantially smaller error bars on the lags compared with our CCF analysis, which may represent an underestimation by JAVELIN, and/or an overestimation by our FR/RSS approach.} The reported 1$\sigma$ uncertainty in the lag measurement gives a rough estimate of the false-positive probability of each detection, but we also perform alternative tests below to estimate the bulk false-positive rate among our reported lags. 

We note that the estimated lag uncertainties are occasionally very asymmetric (e.g., RMID-$320$), reflecting the limitations (e.g., sparse sampling and systematics in flux measurements) of our spectroscopic-only LCs; additional photometric LCs will certainly help in these cases. In addition, although some of these detections appear marginal at best, we keep them in this work as we have adopted rather conservative error bars on the continuum and line LCs, and we expect to improve these measurements by adding photometric LCs that are currently being processed.


To demonstrate that we are mostly detecting real lags instead of false positives based on our spectroscopic-only data, we show in Fig.\ \ref{fig:diag} all the 100 lowest-redshift SDSS-RM quasars in the CCF peak versus line variability signal-to-noise ratio (SNR) plane, again focusing on the \hbeta\ and \MgII\ lines only. The gray points show all the CCF peaks regardless of their significance (searched over a symmetric range of lags centered on zero). The red points show those with a statistical peak significance greater than 0.999, which immediately reveals a preference of positive CCF peaks ($N=55$) over negative ($N=12$) ones (i.e., lags of the lines relative to the continuum, and not vice versa). Objects with larger variability amplitudes in the lines allow more straightforward lag detection (i.e., comparing the gray and red points), as expected. This demonstrates that our spectroscopic-only LCs are able to detect true lags. There is clustering of significant peaks around zero lag, which suggests there are correlated errors in the continuum and line LCs from spectroscopy alone, and/or it reflects the difficulty of detecting lags shorter than the spectroscopic cadence (a few days). Our 15 reported lag detections are shown as red circled points, and they are located in a ``comfortable'' region (i.e., with a large variability amplitude and inconsistent with zero lag) consistent with being genuine lags. There appears to be a deficit of significant peaks around $\sim +10$ days in Fig.\ \ref{fig:diag}, albeit with small number statistics. We suspect this is due to the small gaps between dark/grey time and bright time in our spectroscopic monitoring \citep{Shen_etal_2015a}. If this were the case, we expect the addition of photometric data points in bright time would help recover the lags missed there. On the other hand, our reported lags are all longer than 10 days, and hence should not be significantly affected by these small gaps, as also supported by our tests with JAVELIN (where the LCs are interpolated within these gaps with physically-motivated variability models). 

To further use Fig.\ \ref{fig:diag} as guidance to estimate the bulk false-positive rate among the 15 reported lags, we perform the following tests. 

The first test is a shuffled-epoch test, where we shuffle the LC epochs for each object, and perform the same exercise as in Fig.\ \ref{fig:diag} with real data. By shuffling the LC epochs we destroy any intrinsic correlation, and the frequency of significant CCF peaks from these mock LCs provides one measure of the false-positive rate. We typically find $\sim 4$ cases with a positive peak, a peak significance $>0.999$, and $\rm{S/N_{\rm LC}}>10$, out of 155 LC pairs (\hbeta+\MgII) of the 100 objects. More importantly, we do not observe a preference of positive peaks over negative peaks, as expected from random correlations from temporally uncorrelated data. Given the actual number of positive CCF peaks that pass the statistical significance and ${\rm S/N}$ criteria in real data ($N=40$), this test suggests the bulk false-positive rate is $\sim 10\%$. Note that among the $40$ significant and high S/N positive CCF peaks in the real data, we only selected 15 as our reported detections, and this selection should further reduce false-positives based on visual inspection.

One may be concerned that the shuffled-epoch test does not capture the intrinsic variability characteristics of quasars, and hence may underestimate the false-positive rate. An alternative test is the shuffled-LC-pair test, where we pair the continuum LC of one object to the line LC of another object in our sample, and perform CCF analysis to determine the rate of false positives. This test, however, assumes that individual quasar LCs are sufficiently different that there is no intrinsic correlation in the inter-object LC pairs. In practice, this may not be the case, as different objects may show a similar dominant feature in their LCs (such as a single broad bump/dip) over the limited monitoring period, and thus these inter-object LC pairs will be correlated more often than assumed. Hence the false-positive rate inferred from this test should be treated as an upper limit. Essentially, this test requires a large number of objects to sample the diversity in stochastic quasar variability, in order to justify the basic assumption that these inter-object LC pairs are intrinsically uncorrelated. 

We performed the shuffled-LC-pair test for all 100 objects, focusing on \hbeta\ and \MgII\ only (15336 inter-object LC pairs). We found roughly equal numbers of positive and negative peaks (581 versus 523) with a peak significance $>0.999$ and $\rm{S/N_{\rm LC}}>10$, confirming the results based on the shuffled-epoch test. We also found that $3.8\%$ of these inter-object LC pairs lead to a positive CCF peak with $p>0.999$ and $\rm{S/N_{\rm LC}}>10$. Given the number of real LC pairs (155) in the parent sample, we therefore estimate that $\sim 6$ cases will manifest from uncorrelated LCs as passing the S/N and the peak significance criteria as for the real data. This is about $15\%$ of the $40$ cases we observed with real data, which is an upper limit as explained above. 

Therefore based on the shuffled-epoch test and the shuffled-LC-pair test, we consistently estimate a false-positive rate of $\sim 10-15\%$ among our reported lags. This false-positive rate is also roughly consistent with the estimation from our simulations using mock quasar LC data and the SDSS-RM sampling/spectral quality \citep[][]{Shen_etal_2015a}. 


The basic properties of these lags are summarized in Table~\ref{table:basic}, and the spectroscopic light curves are provided in Table \ref{table:lc}. There are 9 \hbeta\ lags and 6 \MgII\ lags. As mentioned in \S\ref{sec:intro}, we do not consider multiple line detections in the same object in the current work, but simply note that there are cases where more than one line have detected (and consistent) lags, while in other cases there is only one line detected, and the other lines show signs of a lag but will require further data and in-depth analysis (see \S\ref{sec:con}). The important aspects of multiple line detections in the same object and their implications for BLR structure will be the focus of successive publications. 

Epochs 3 and 7 have much lower S/N than the other epochs in our data \citep[see Table 2 of][]{Shen_etal_2015a}. For these two epochs, the narrow line emission is often too noisy to use as a reliable internal flux calibrator, and as a result the data points at these two epochs sometimes appear as significant outliers (albeit with large error bars) from the adjacent LC data points. The standard interpolated CCF method does not treat the errors in the LCs rigorously and may be significantly affected by these discrepant or noisy measurements. Therefore in such cases, we remove these problematic epochs (primarily Epoch 7, sometimes both Epoch 7 and 3), and rely on the rest of the epochs for the lag measurement. We only reject these epochs when necessary (i.e., if they significantly degrade the CCF), as rejecting data in general will reduce the correlation by reducing the sampling of the light curves. These rejected epochs are identified with a ``1'' in the last column in Table \ref{table:lc}. 

\begin{figure*}
\centering
    \includegraphics[width=0.48\textwidth]{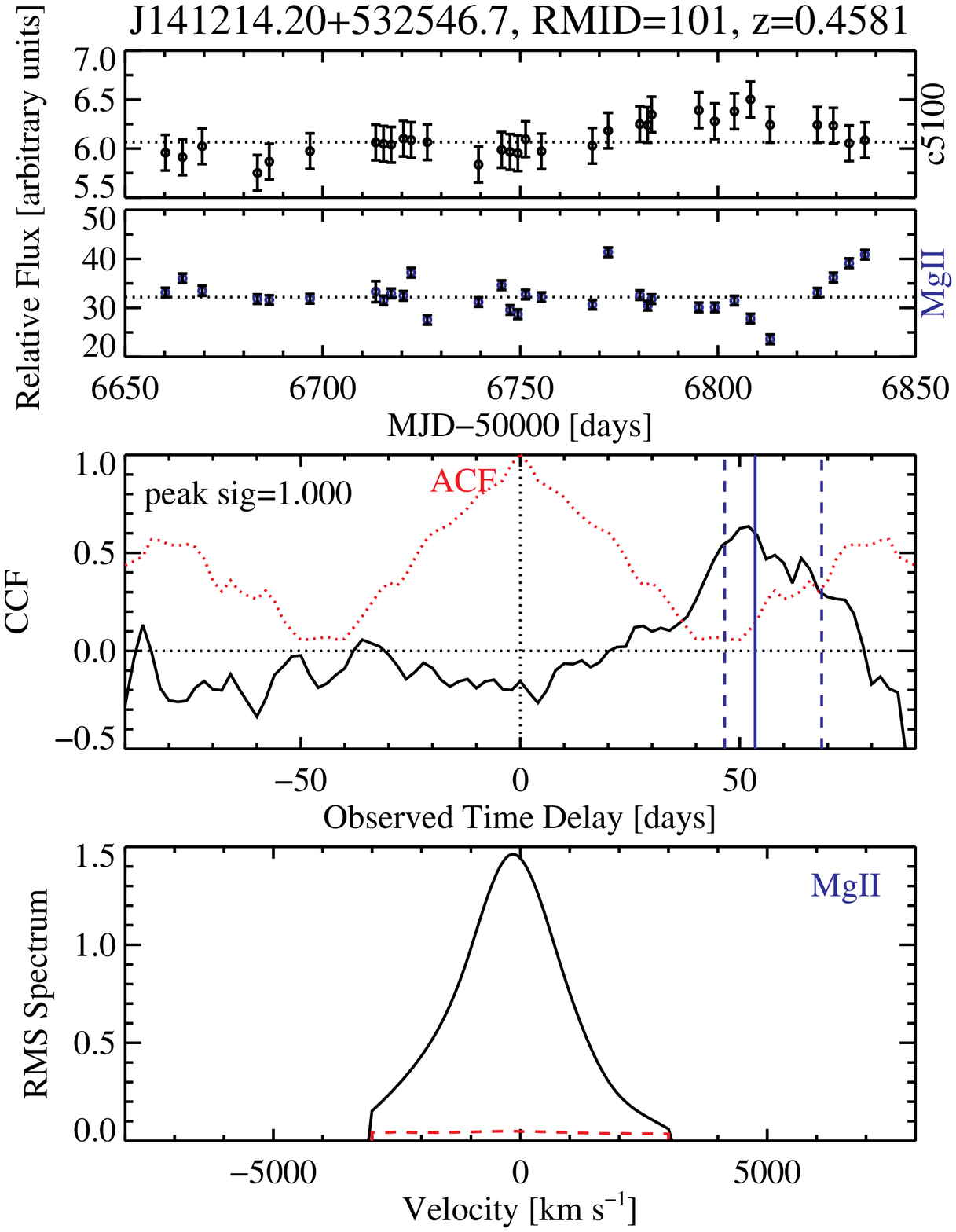}\vspace{3mm}
    \includegraphics[width=0.48\textwidth]{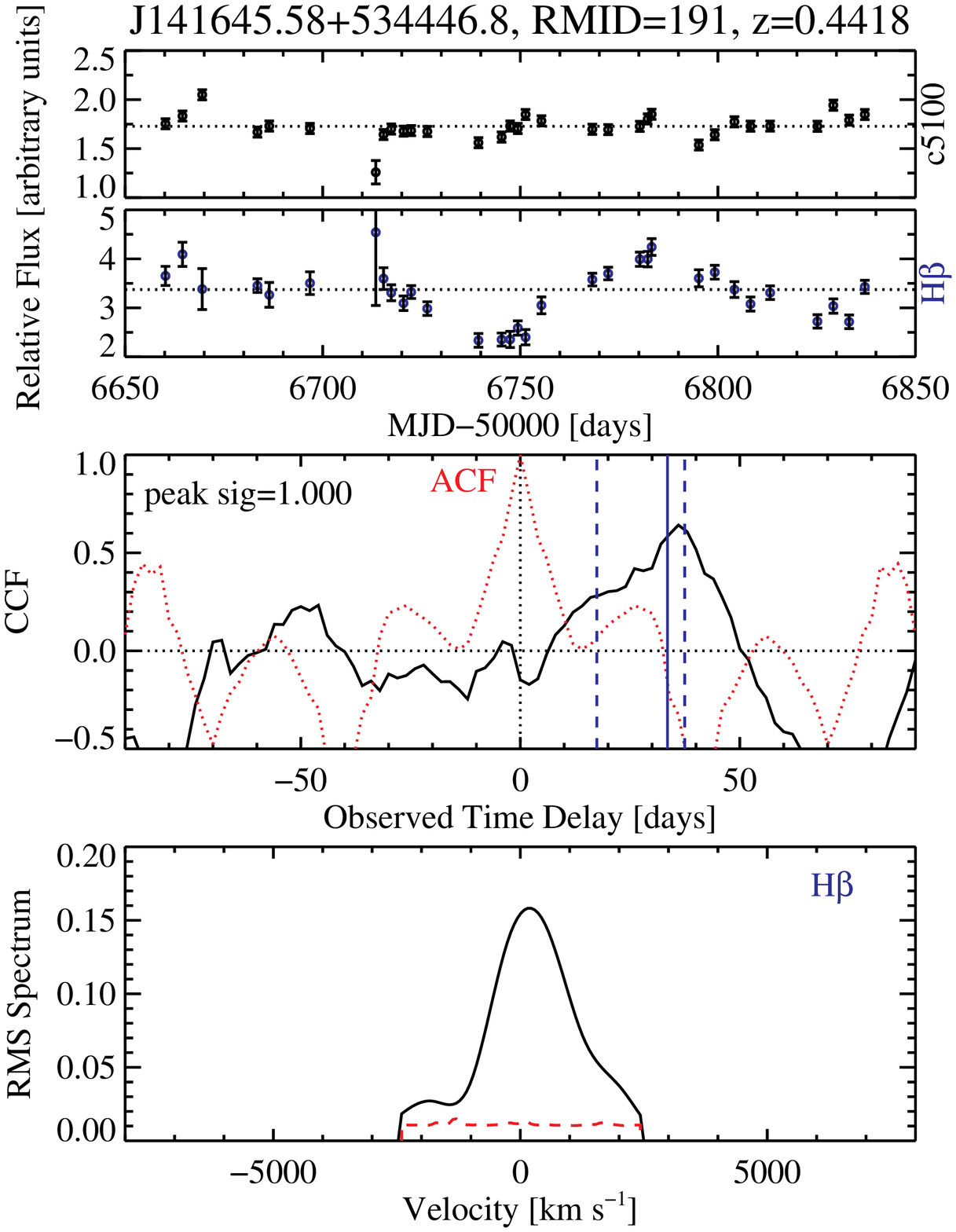}\vspace{3mm}
    \includegraphics[width=0.48\textwidth]{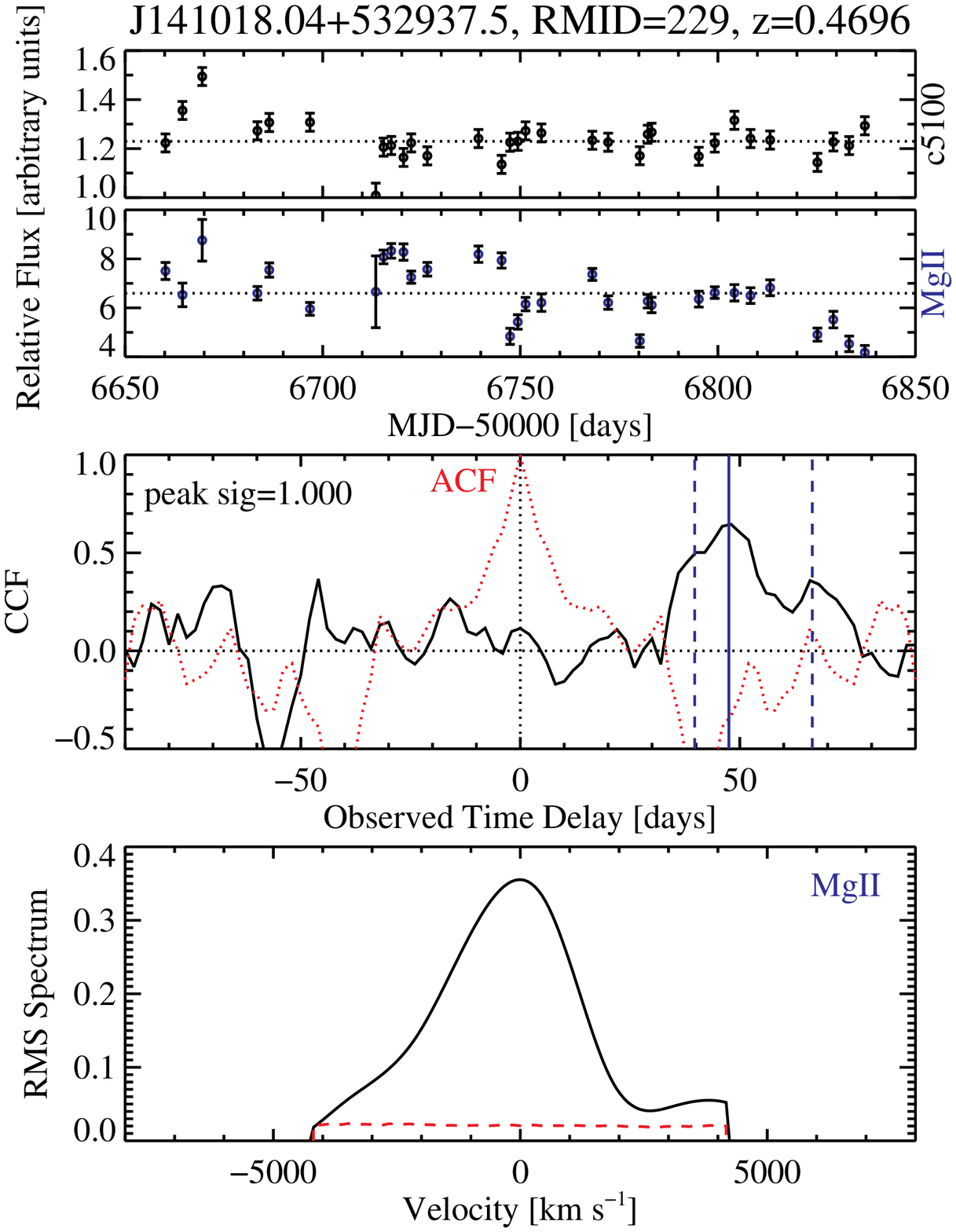}
    \includegraphics[width=0.48\textwidth]{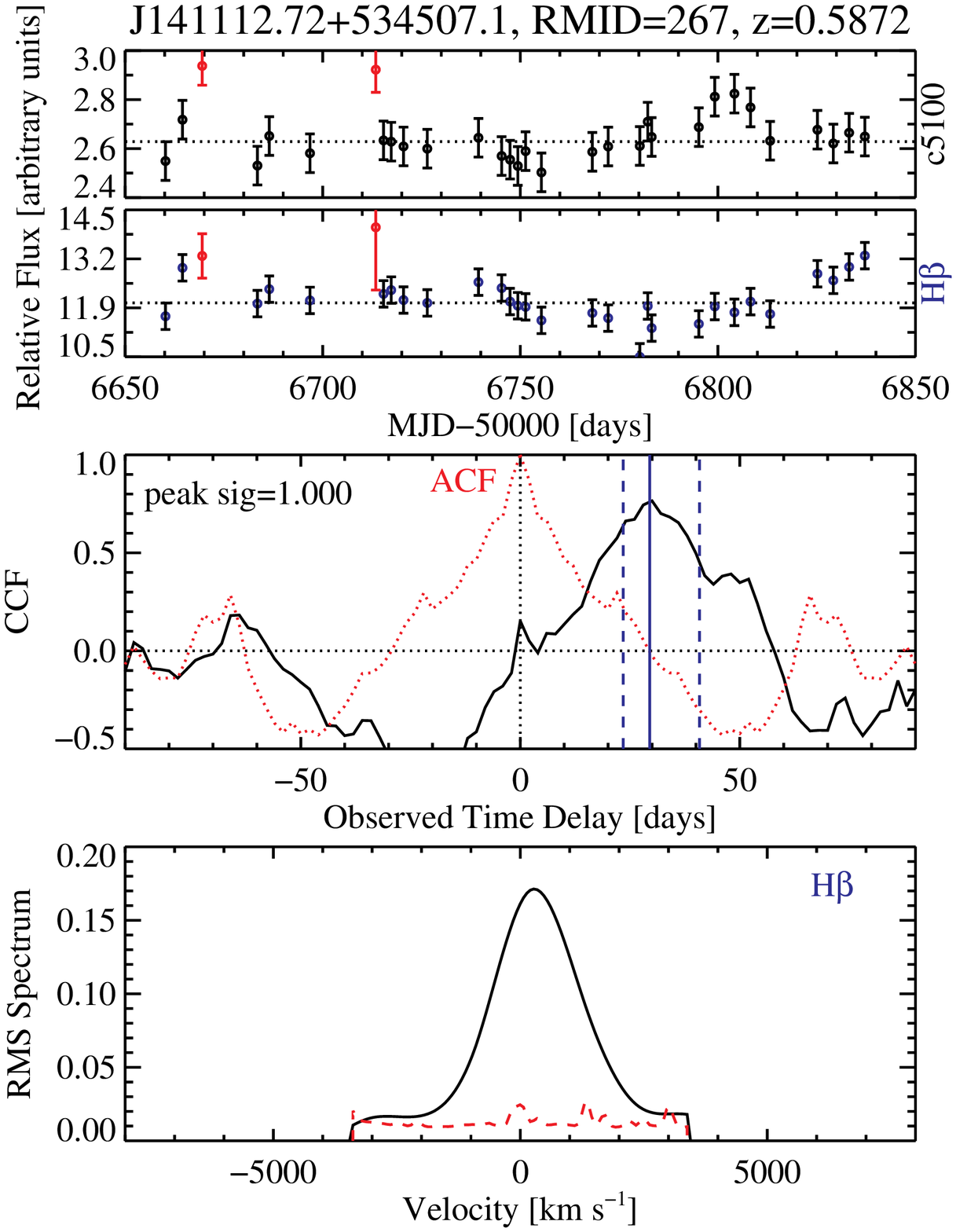}
     \caption{Light curves, CCF, and RMS line profile for the 15 objects with lag detections. For each detection, the top panel shows the continuum (at rest-frame 5100\,\AA) and broad-line light curves, with the median flux indicated by the dotted horizontal line. Bad epochs are marked in red and excluded from the CCF analysis (see text for details). The middle panel shows the CCF (solid black line), and the auto correlation function (ACF) of the continuum LC is shown in the red dotted line. The lag (i.e., the median of the CCF centroid distribution from FR/RSS; see \S\ref{sec:lag_det}) is indicated by the solid vertical line, and the dashed vertical lines indicate the 1$\sigma$ uncertainty in the lag. The statistical significance of the CCF peak is shown in the upper-left corner. The bottom panel shows the model RMS broad-line flux in the black line and the estimated errors in the red dashed line, both output by PrepSpec. We only show RMS flux errors within the adaptive broad-line fitting window, as errors outside the fitting window are not properly estimated in PrepSpec. }
    \label{fig:ccf}
\end{figure*}

\begin{figure*}
\centering
    \includegraphics[width=0.48\textwidth]{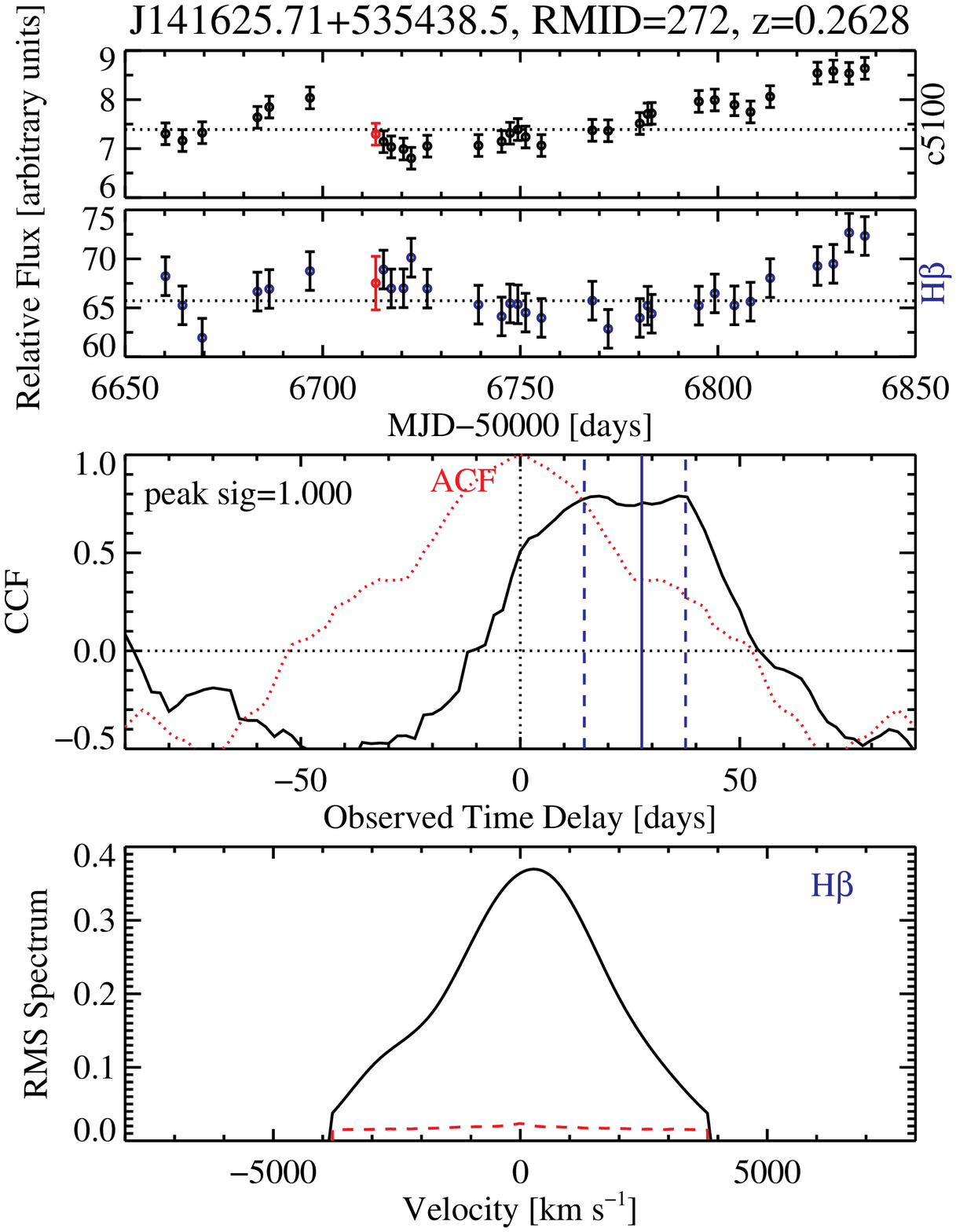}\vspace{3mm}
    \includegraphics[width=0.48\textwidth]{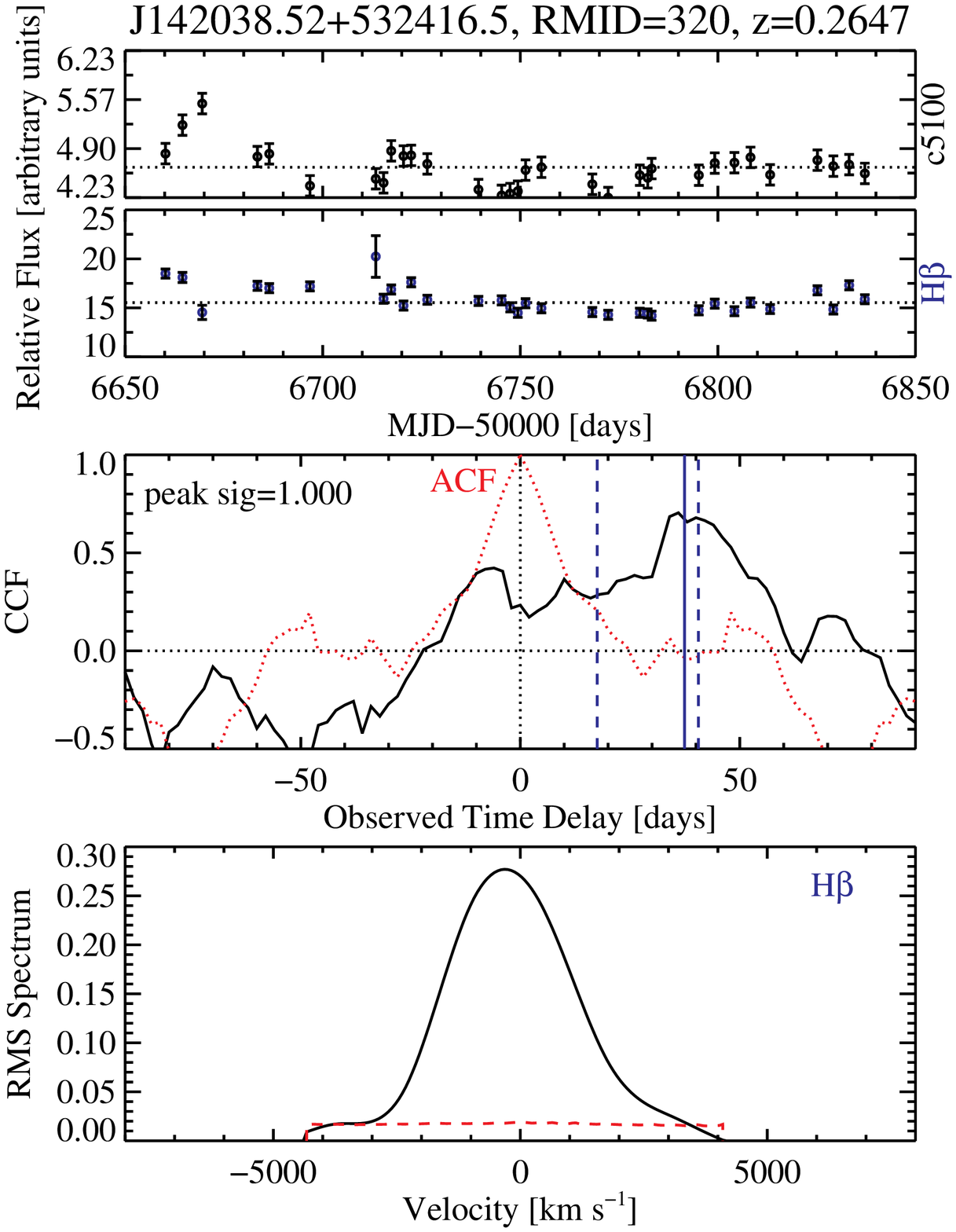}\vspace{3mm}
    \includegraphics[width=0.48\textwidth]{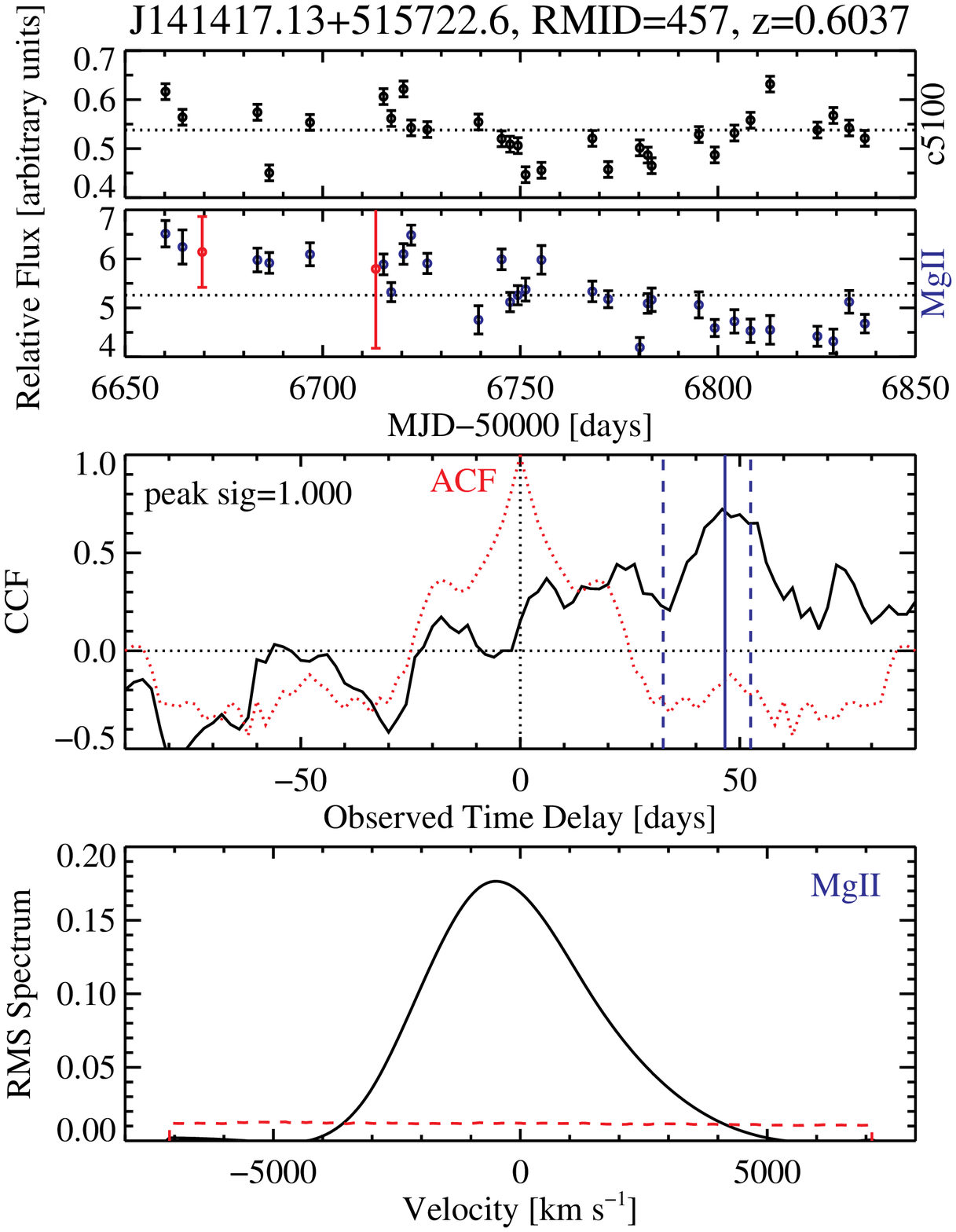}
    \includegraphics[width=0.48\textwidth]{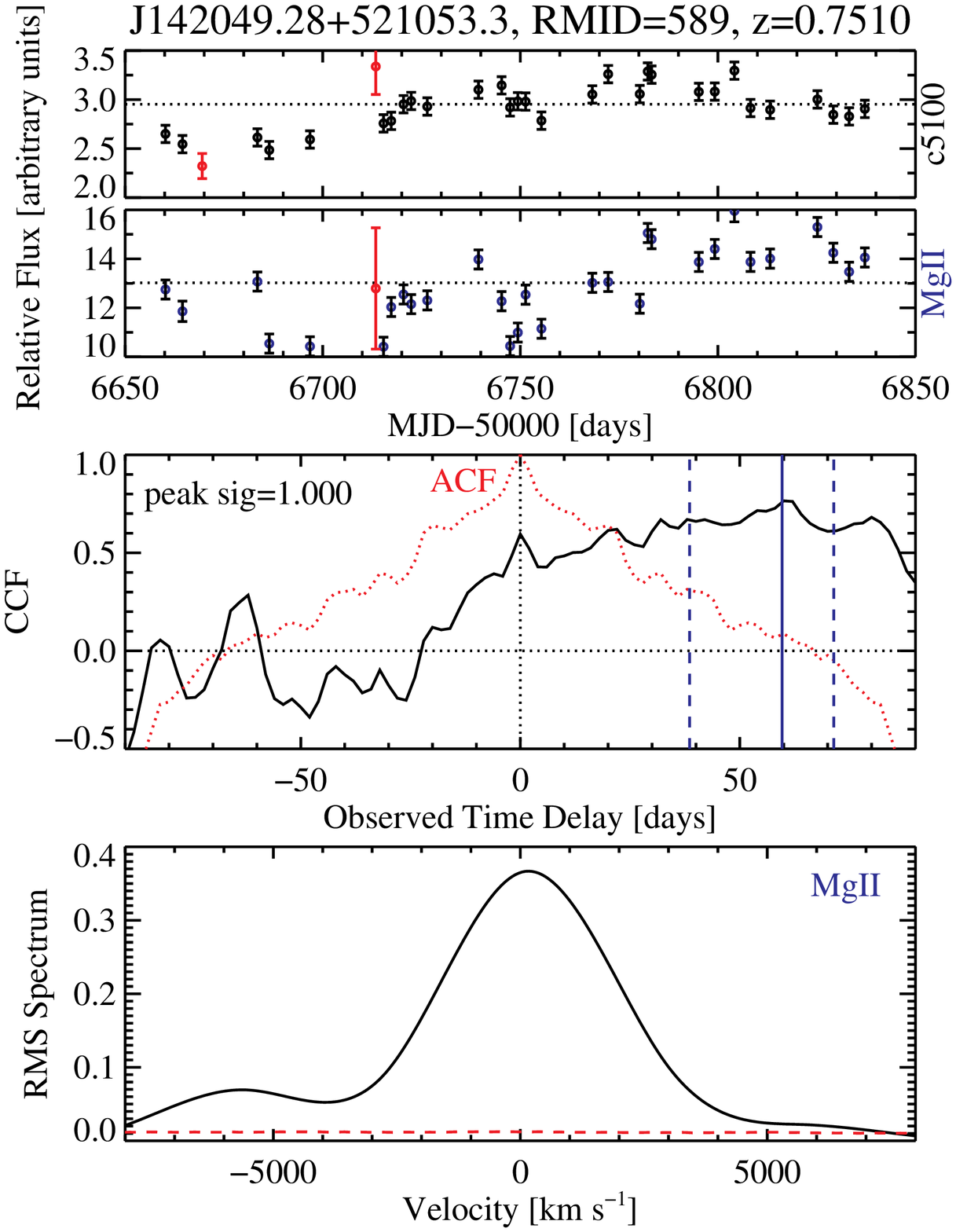}
     \caption{Same as Fig.\ \ref{fig:ccf}, for another set of 4 objects with lag measurements.}
    \label{fig:ccf2}
\end{figure*}

\begin{figure*}
\centering
    \includegraphics[width=0.48\textwidth]{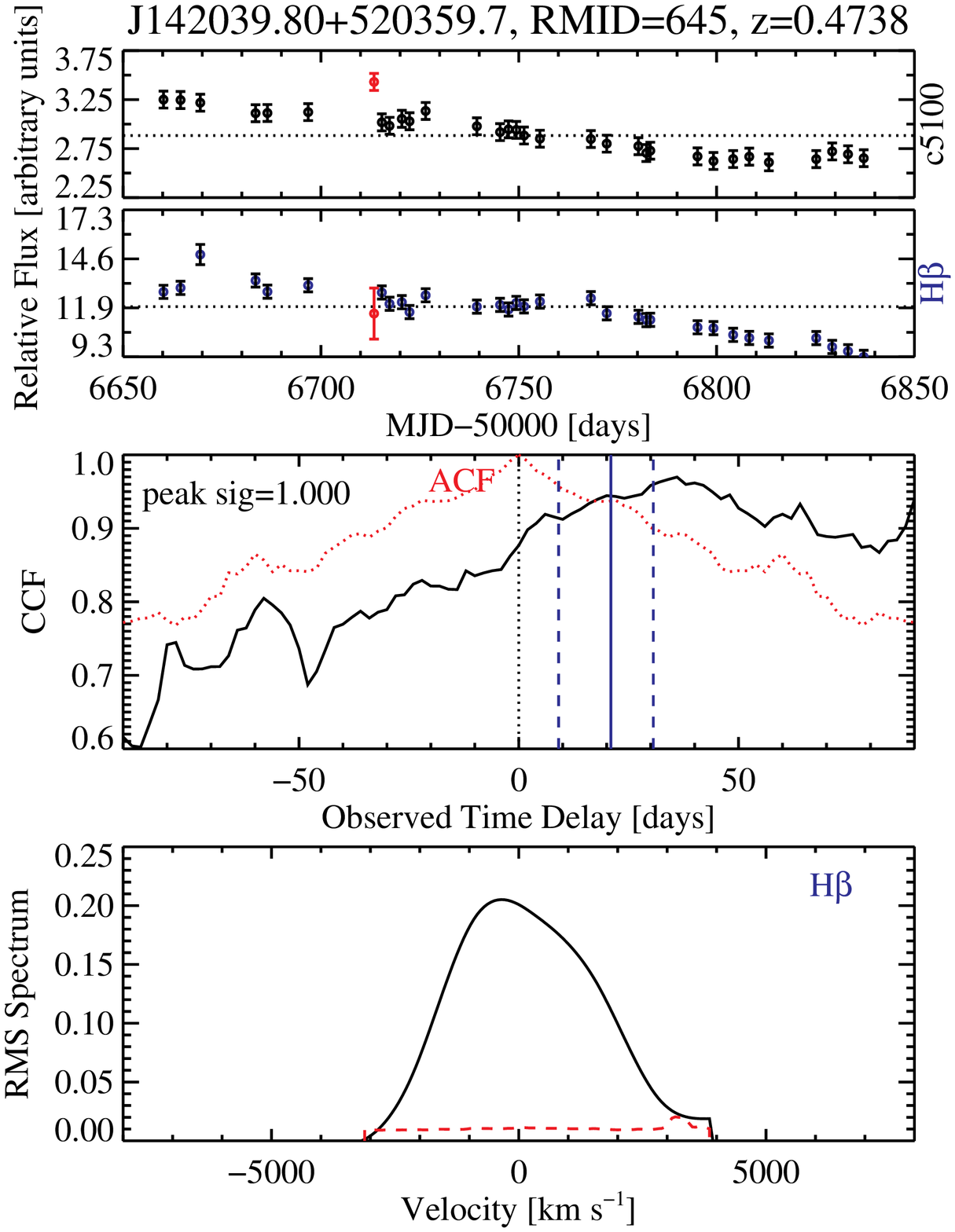}\vspace{3mm}
    \includegraphics[width=0.48\textwidth]{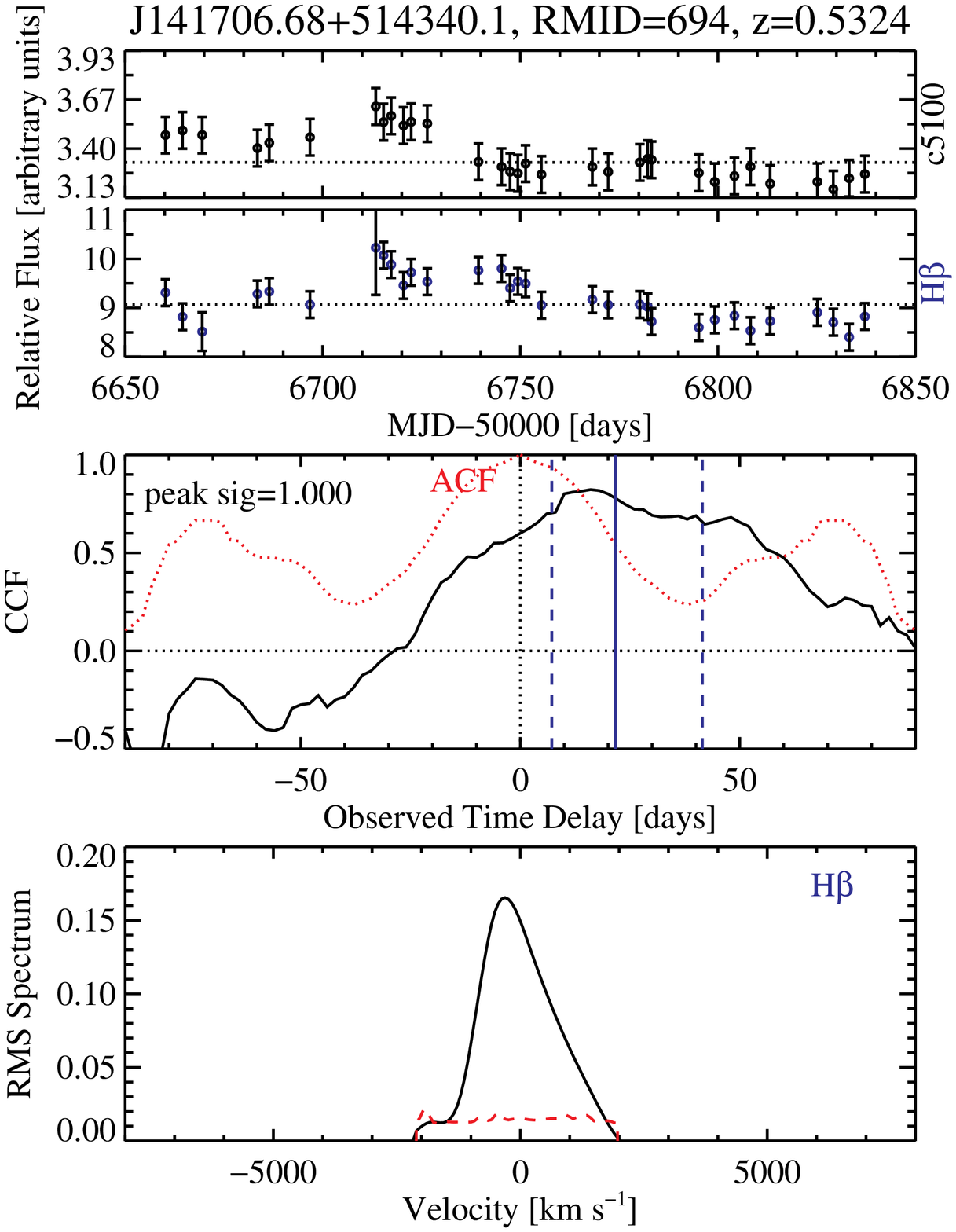}\vspace{3mm}
    \includegraphics[width=0.48\textwidth]{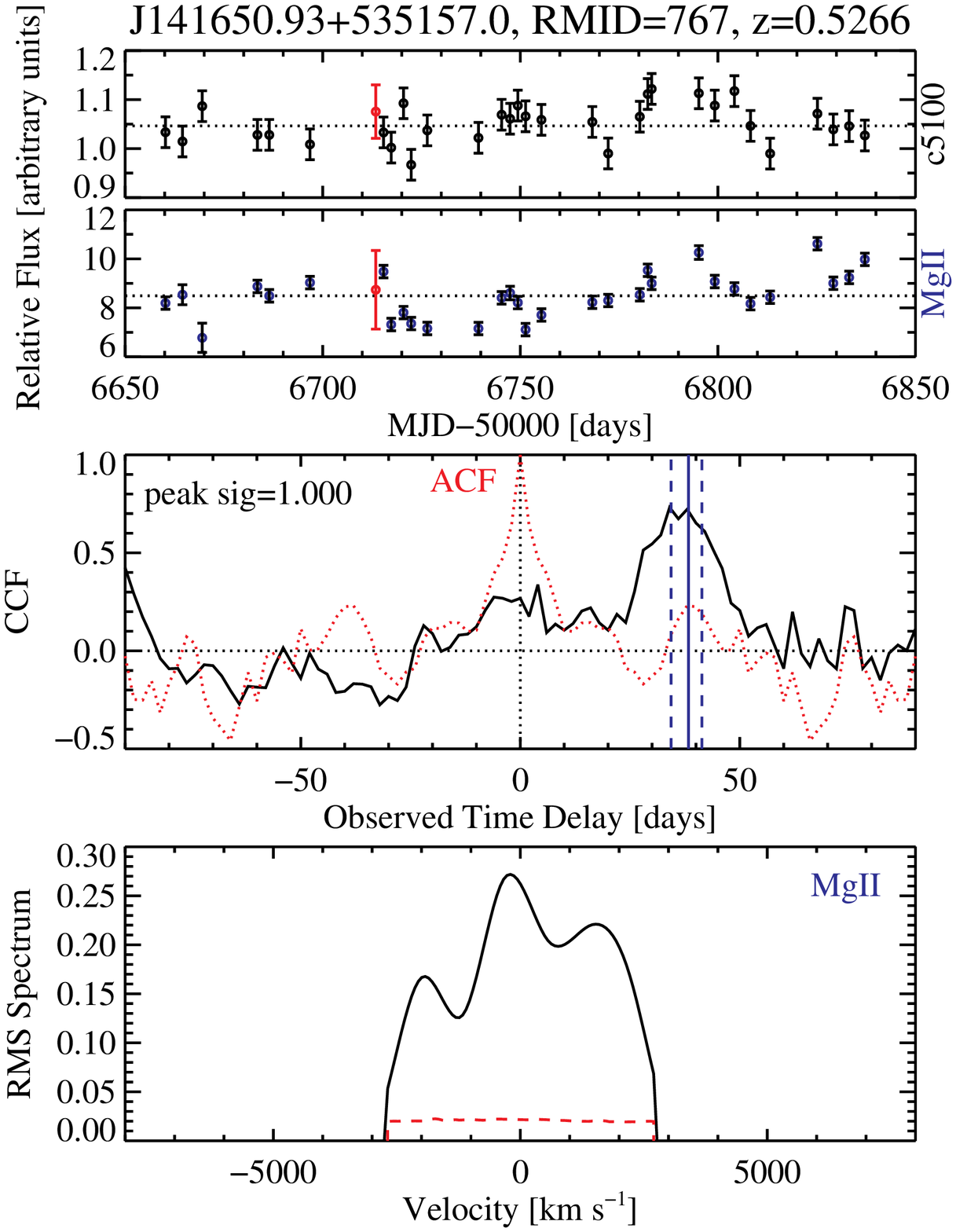}
    \includegraphics[width=0.48\textwidth]{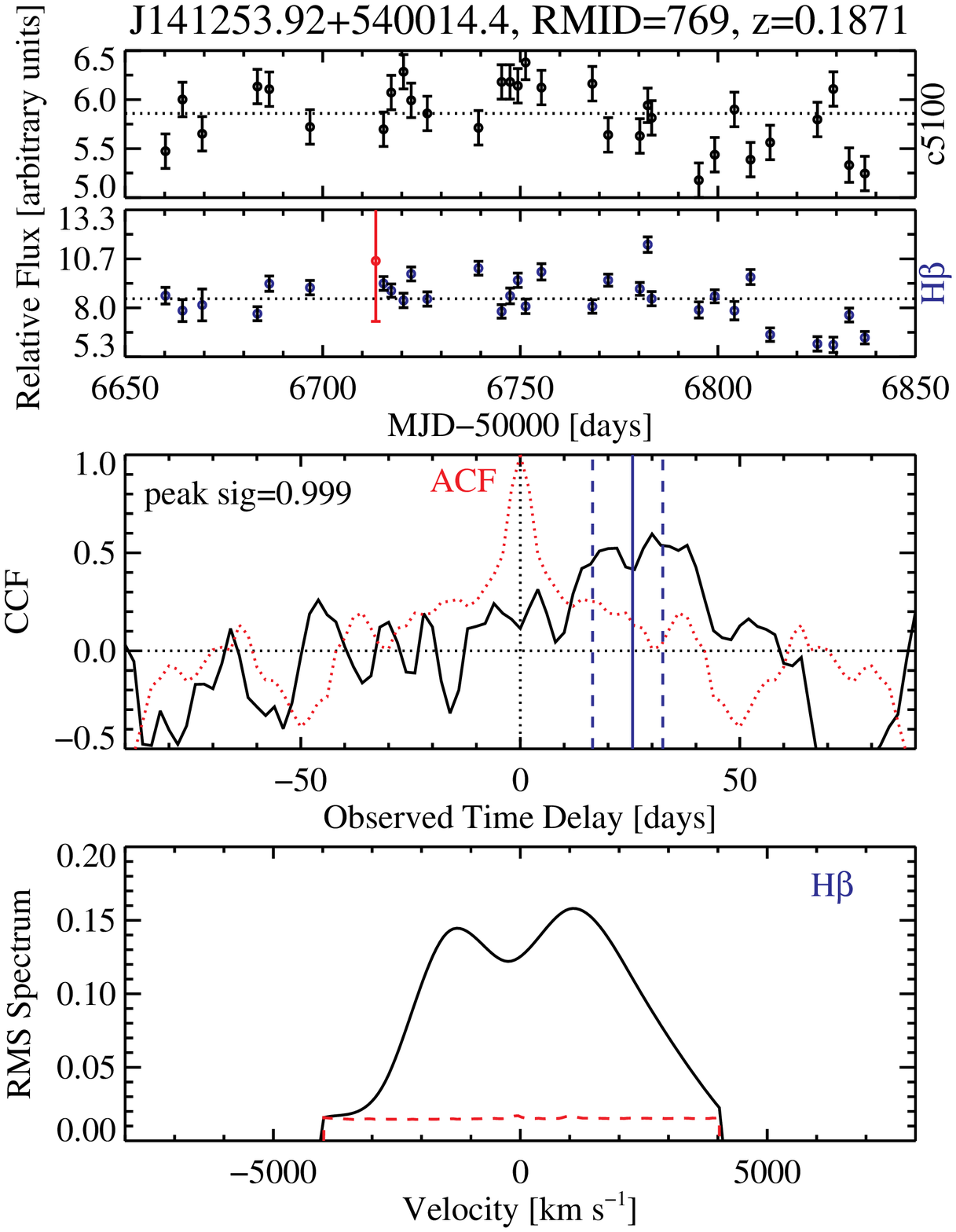}
     \caption{Same as Fig.\ \ref{fig:ccf}, for another set of 4 objects with lag measurements. }
    \label{fig:ccf3}
\end{figure*}

\begin{figure*}
\centering
    \includegraphics[width=0.48\textwidth]{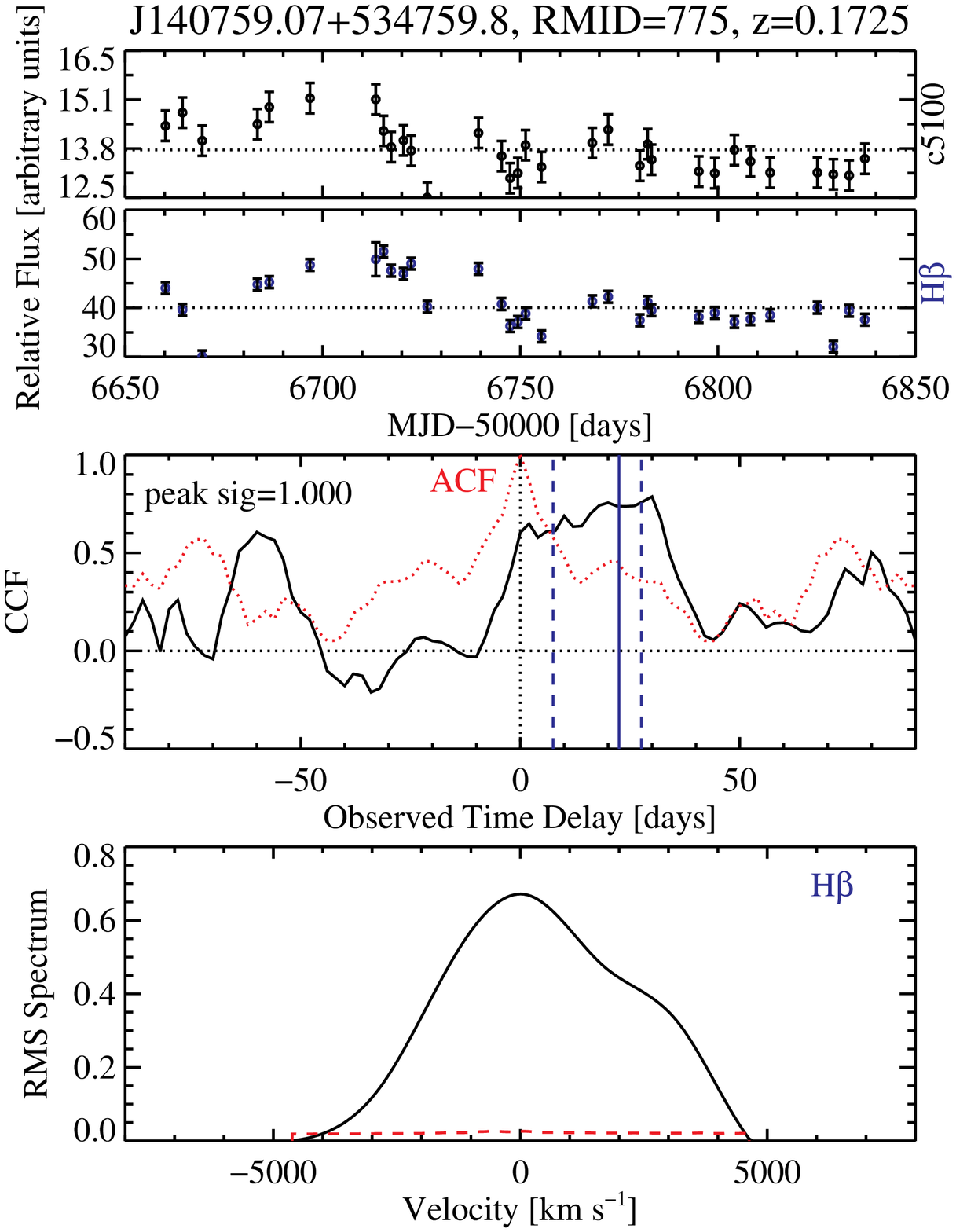}\vspace{3mm}
    \includegraphics[width=0.48\textwidth]{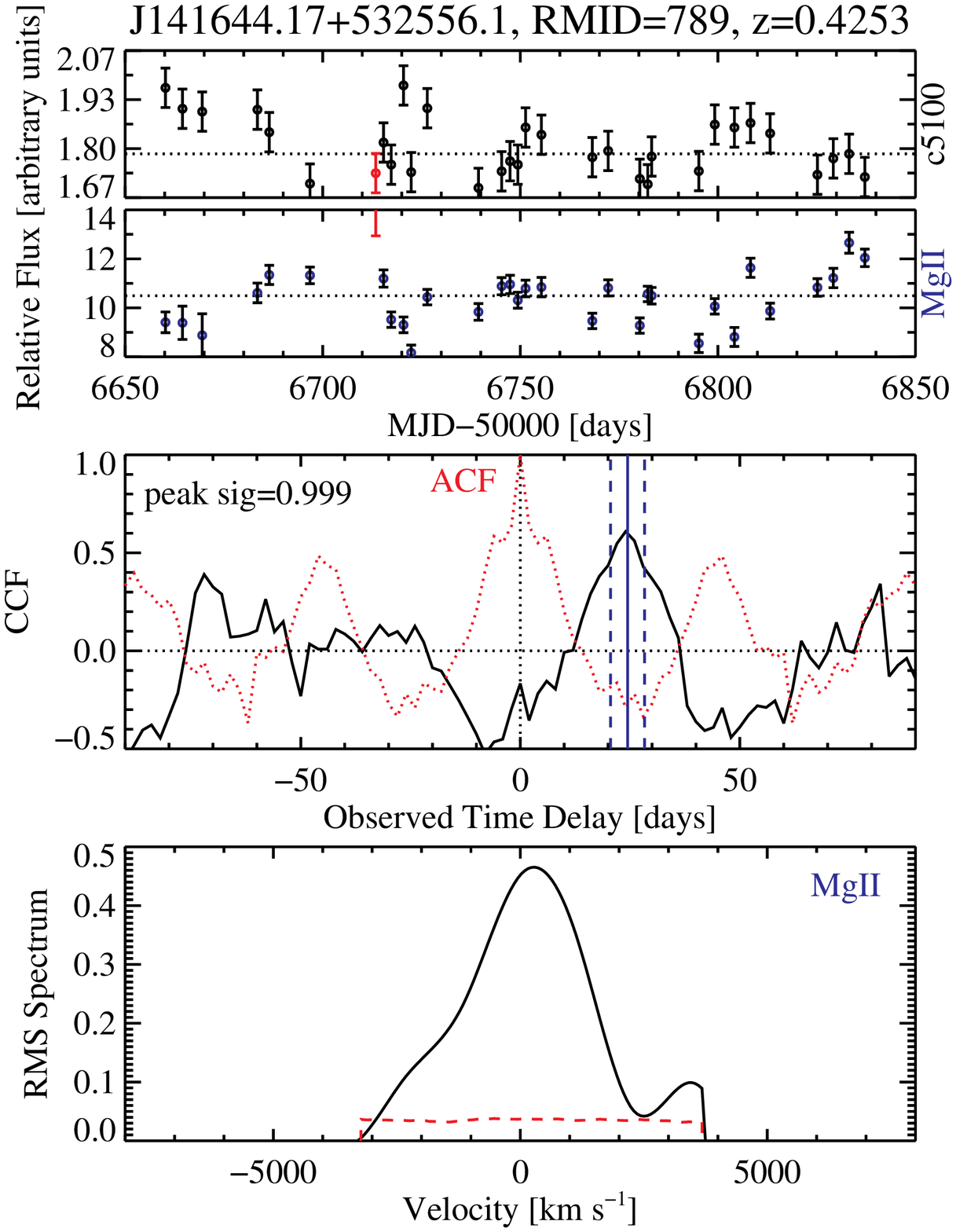}\vspace{3mm}
    \includegraphics[width=0.48\textwidth]{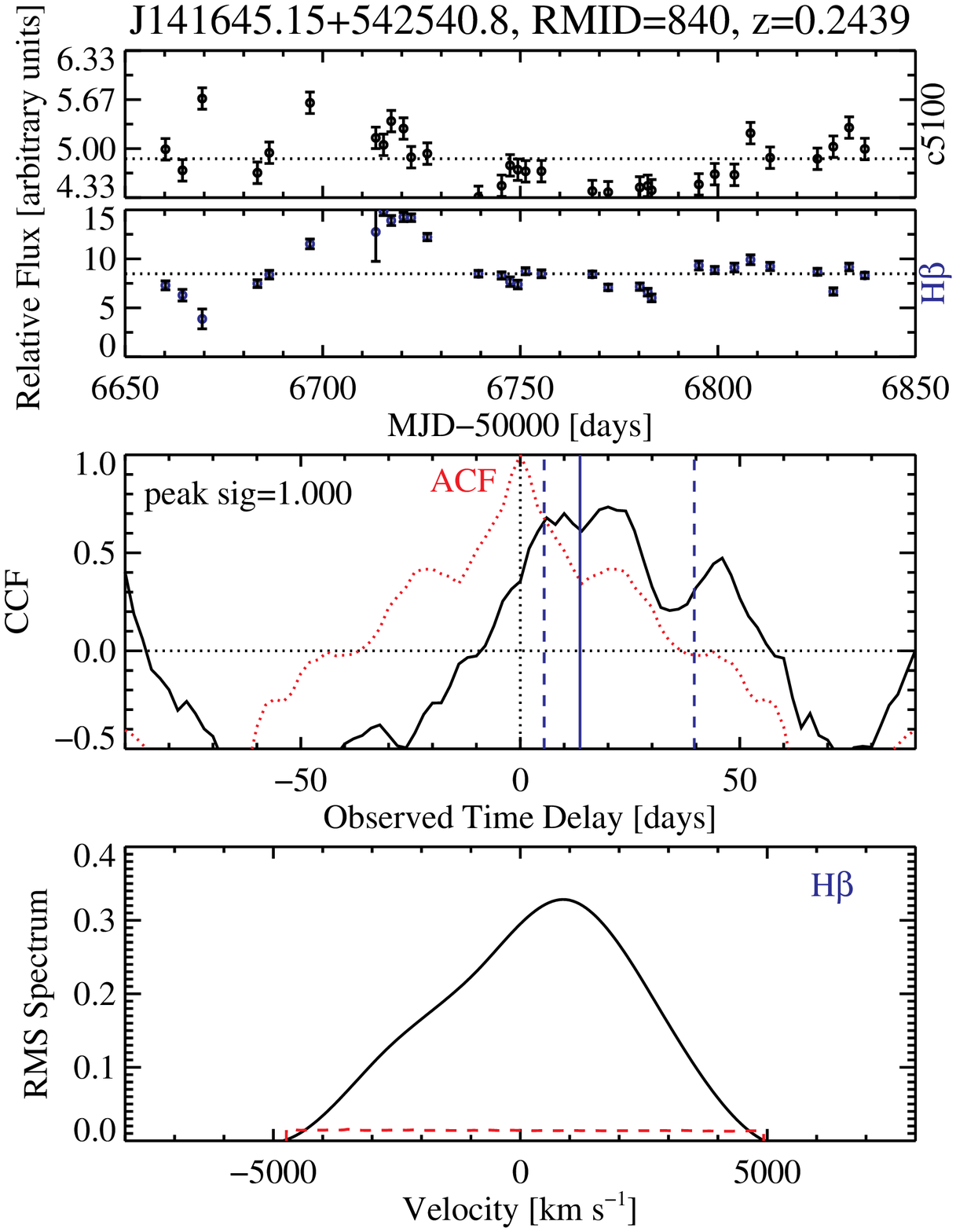}
     \caption{Same as Fig.\ \ref{fig:ccf}, for another set of 3 objects with lag measurements. }
    \label{fig:ccf4}
\end{figure*}

\begin{figure}
\centering
    \includegraphics[width=0.48\textwidth]{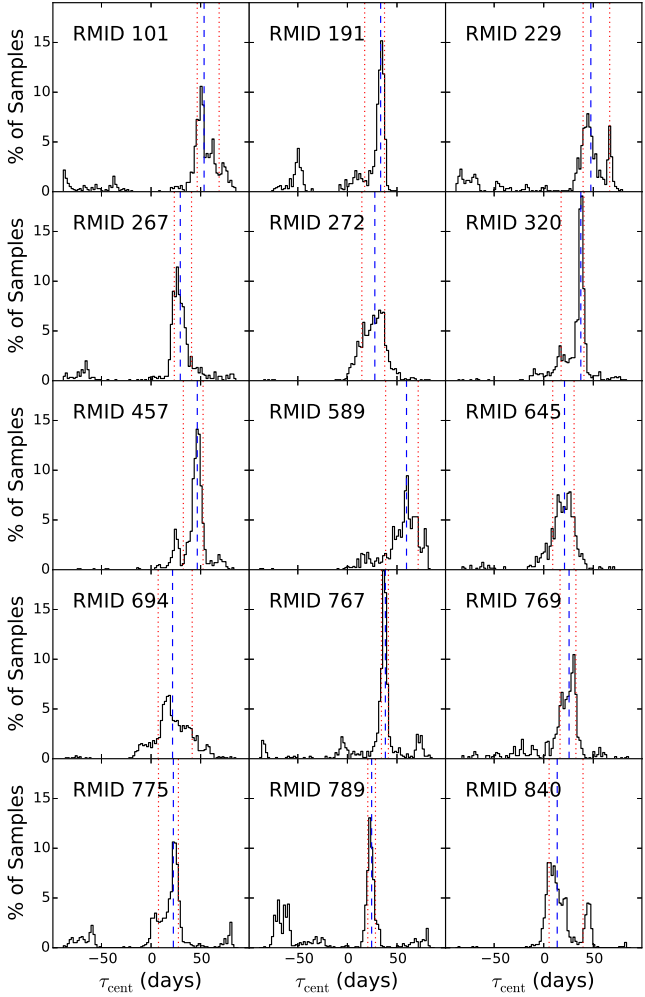}
     \caption{{The CCF centroid distributions (CCCDs) from FR/RSS for the 15 lags. The lags are in the observed frame to match Figs.\ \ref{fig:ccf}--\ref{fig:ccf4}. The vertical dashed and dotted lines indicate the reported lag and its uncertainties. In all cases there is a reasonably well-defined main peak in the CCCD to determine the best lag. In a few cases there are sub-structures (possible aliases due to the sparse sampling of the LCs) in the CCCD that will lead to elevated uncertainties in the lag. }}
    \label{fig:cccd}
\end{figure}

\begin{figure}
\centering
    \includegraphics[width=0.48\textwidth]{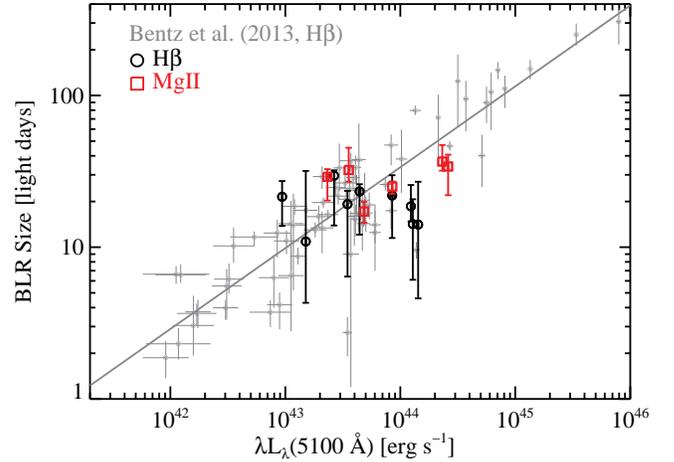}
     \caption{The BLR size-luminosity relation. Our lag detections are shown as black circles (for \hbeta\ detections) and red squares (for \MgII\ detections). The data for previous $z<0.3$ RM AGN compiled in \citet{Bentz_etal_2013} are indicated in gray points. Our new lags are consistent with the locations of the previous RM AGN used to calibrate the local $R-L$ relation, but are not yet able to constrain the $R-L$ relation independently given the limited numbers, precision, dynamic range, and possible selection biases inherent to our program (see discussion in the text). }
    \label{fig:rl_relation}
\end{figure}



\begin{table*}
\caption{Basic properties of the lag detections}\label{table:basic}
\centering
\scalebox{0.7}{
\begin{tabular}{lcccccccccccccc}
\hline\hline
RMID & SDSS designation & Redshift & Morphology & Line & $\sigma_{\rm rms,line}$ & $\tau_{\rm cent}$ & $\log {\rm VP}$ & $f_{5100}$ & $\log L_{5100}$ & $f_{\rm host}$ & FWHM$_{\rm mean,H\beta}$ &  $\log M_{\rm SE,H\beta}$ & $\sigma_{\rm mean,line}$ & FWHM$_{\rm rms,line}$ \\
& hhmmss.ss$\pm$ddmmss.s & & & & $(\kms)$ & (days) & [$M_\odot$] & ($10^{-17}\,{\rm erg\,s^{-1}cm^{-2}\textrm{\AA}^{-1}}$) & [${\rm erg\,s^{-1}}$] & & $(\kms)$ & [$M_\odot$] & $(\kms)$ & $(\kms)$ \\                                                                                      
(1) & (2) & (3) & (4) & (5) & (6) & (7) & (8) & (9) & (10) & (11) & (12) & (13) & (14) & (15)\\                                                                                                                                                                                    
\hline
101 &  141214.20$+$532546.7 &  0.4581 &     point &  \MgII  & $1167\pm  9$ & $36.7_{ -4.8}^{+10.4}$ & $6.989_{-0.057}^{+0.123}$ & $3.998\pm0.005$ & $44.365\pm0.001$ & 0.09 & $2443\pm  12$ & $7.889\pm0.004$ & $1029\pm  4$ & $2391\pm  46$ \\ 
191 &  141645.58$+$534446.8 &  0.4418 &  extended &  \hbeta & $ 990\pm 19$ & $23.3_{-11.2}^{+ 2.7}$ & $6.648_{-0.209}^{+0.053}$ & $0.842\pm0.015$ & $43.646\pm0.008$ & 0.48 & $2173\pm  34$ & $7.550\pm0.014$ & $ 838\pm 12$ & $1854\pm  70$ \\ 
229 &  141018.04$+$532937.5 &  0.4696 &  extended &  \MgII  & $1630\pm 24$ & $32.3_{ -5.3}^{+12.9}$ & $7.225_{-0.072}^{+0.174}$ & $0.575\pm0.004$ & $43.551\pm0.003$ & 0.49 & $3854\pm 331$ & $8.003\pm0.075$ & $1582\pm 20$ & $3101\pm  76$ \\ 
267 &  141112.72$+$534507.1 &  0.5872 &     point &  \hbeta & $1221\pm 36$ & $18.6_{ -3.8}^{+ 7.1}$ & $6.733_{-0.092}^{+0.169}$ & $1.078\pm0.004$ & $44.092\pm0.002$ & 0.39 & $2680\pm  61$ & $7.921\pm0.020$ & $1403\pm  6$ & $2089\pm  77$ \\ 
272 &  141625.71$+$535438.5 &  0.2628 &     point &  \hbeta & $1636\pm 11$ & $21.9_{-10.4}^{+ 7.9}$ & $7.059_{-0.205}^{+0.157}$ & $6.190\pm0.254$ & $43.929\pm0.018$ & 0.00 & $2983\pm  51$ & $7.824\pm0.017$ & $1515\pm  2$ & $3752\pm  93$ \\ 
320 &  142038.52$+$532416.5 &  0.2647 &  extended &  \hbeta & $1362\pm 33$ & $29.6_{-15.7}^{+ 2.5}$ & $7.030_{-0.232}^{+0.042}$ & $1.900\pm0.007$ & $43.424\pm0.001$ & 0.47 & $4462\pm 114$ & $8.057\pm0.022$ & $1576\pm  9$ & $2975\pm  64$ \\ 
457 &  141417.13$+$515722.6 &  0.6037 &     point &  \MgII  & $1672\pm 60$ & $29.1_{ -8.8}^{+ 3.6}$ & $7.201_{-0.135}^{+0.063}$ & $0.188\pm0.005$ & $43.366\pm0.012$ & 0.60 & $4505\pm 541$ & $8.101\pm0.105$ & $2574\pm 39$ & $3874\pm  86$ \\ 
589 &  142049.28$+$521053.3 &  0.7510 &     point &  \MgII  & $2824\pm 33$ & $34.0_{-12.0}^{+ 6.7}$ & $7.724_{-0.154}^{+0.086}$ & $1.132\pm0.006$ & $44.416\pm0.002$ & 0.20 & $4750\pm 117$ & $8.521\pm0.021$ & $3283\pm 22$ & $4108\pm  39$ \\ 
645 &  142039.80$+$520359.7 &  0.4738 &     point &  \hbeta & $1360\pm 20$ & $14.2_{ -8.1}^{+ 6.5}$ & $6.711_{-0.247}^{+0.200}$ & $2.026\pm0.004$ & $44.109\pm0.001$ & 0.12 & $4122\pm  58$ & $8.222\pm0.012$ & $1571\pm  7$ & $3696\pm  55$ \\ 
694 &  141706.68$+$514340.1 &  0.5324 &     point &  \hbeta & $ 743\pm 24$ & $14.1_{ -9.5}^{+12.9}$ & $6.183_{-0.292}^{+0.398}$ & $1.635\pm0.004$ & $44.155\pm0.001$ & 0.23 & $1888\pm  18$ & $7.595\pm0.008$ & $ 864\pm  3$ & $1661\pm 104$ \\ 
767 &  141650.93$+$535157.0 &  0.5266 &  extended &  \MgII  & $1394\pm 10$ & $25.1_{ -2.6}^{+ 2.0}$ & $6.979_{-0.045}^{+0.035}$ & $1.004\pm0.002$ & $43.930\pm0.001$ & 0.00 & $2088\pm  94$ & $7.514\pm0.039$ & $1341\pm  4$ & $4066\pm 202$ \\ 
769 &  141253.92$+$540014.4 &  0.1871 &  extended &  \hbeta & $1758\pm 22$ & $21.5_{ -7.7}^{+ 5.8}$ & $7.114_{-0.155}^{+0.117}$ & $1.563\pm0.010$ & $42.972\pm0.003$ & 0.72 & $4192\pm 109$ & $7.920\pm0.023$ & $1769\pm 13$ & $5120\pm 130$ \\ 
775 &  140759.07$+$534759.8 &  0.1725 &  extended &  \hbeta & $1790\pm 10$ & $19.2_{-12.8}^{+ 4.3}$ & $7.079_{-0.290}^{+0.098}$ & $7.023\pm0.008$ & $43.541\pm0.001$ & 0.44 & $3661\pm  33$ & $7.933\pm0.008$ & $1615\pm  5$ & $5115\pm  59$ \\ 
789 &  141644.17$+$532556.1 &  0.4253 &  extended &  \MgII  & $1371\pm 27$ & $17.2_{ -2.7}^{+ 2.7}$ & $6.799_{-0.071}^{+0.071}$ & $1.020\pm0.005$ & $43.685\pm0.002$ & 0.27 & $4128\pm 138$ & $8.052\pm0.029$ & $1328\pm 13$ & $2681\pm  96$ \\ 
840 &  141645.15$+$542540.8 &  0.2439 &  extended &  \hbeta & $1902\pm 20$ & $10.9_{ -6.6}^{+20.9}$ & $6.888_{-0.261}^{+0.832}$ & $1.321\pm0.007$ & $43.178\pm0.002$ & 0.67 & $5923\pm 191$ & $8.287\pm0.028$ & $2492\pm 29$ & $4981\pm  97$ \\ 
\hline
\hline\\
\end{tabular}
}
\begin{tablenotes}
      \small
      \item NOTE. --- Col (1): object index in the full SDSS-RM sample described in \citet{Shen_etal_2015a}. Col (2): object designation in J2000 coordinates. Col (3): redshift. Col (4): object morphological classification based on SDSS imaging. Col (5): the broad line used for the lag detection. Col (6): broad-line dispersion (second moment) for the line specified in Col (5) measured by PrepSpec using the RMS spectrum. Col (7): rest-frame time lag from the centroid of the CCF peak. Col (8): virial product defined in Eqn.\ (\ref{eqn:vp}), which can be converted to the RM-based BH mass $\log M_{\rm RM}=\log {\rm VP} + \log f$ with $f=5.5$ adopted in this work. Col (9): observed quasar continuum flux (host-corrected) at rest-frame 5100\,\AA, measured from spectral fits to the mean spectrum. Col (10): quasar continuum luminosity (host-corrected) at rest-frame 5100\,\AA. Col (11): host to total fraction in 5100\,\AA\ continuum luminosity, estimated using a spectral decomposition approach \citep[][]{Shen_etal_2015b}. Col (12): \hbeta\ broad-line FWHM measured from the mean spectrum. Col (13): single-epoch BH mass estimate based on the \hbeta\ FWHM and continuum luminosity measured from the mean spectrum, using the formula from \citet{Vestergaard_Peterson_2006}. Col (14): broad-line dispersion for the line specified in Col (5) measured by PrepSpec using the mean spectrum. Col (15): broad-line FWHM for the line specified in Col (5) measured by PrepSpec using the RMS spectrum. All uncertainties presented here are 1$\sigma$ statistical errors only. 
\end{tablenotes}
\end{table*}

\begin{table}
\caption{Continuum and broad-line light curves}\label{table:lc}
\centering
\scalebox{0.9}{
\begin{tabular}{lcccccc}
\hline\hline
RMID & MJD & $f_{\rm cont}$ & $e_{\rm cont}$ & $f_{\rm line}$ & $e_{\rm line}$ & Mask \\
\hline
101 & 56660.209 & 5.959 & 0.008 & 33.151 &  0.511 & 0 \\
\hline
\hline\\
\end{tabular}
}
\begin{tablenotes}
      \small
      \item NOTE. --- Continuum and broad-line light curves (time series of fluxes) are simultaneously derived from spectroscopy using PrepSpec (\S\ref{sec:prepspec}), and the continuum is always estimated at rest-frame 5100\,\AA. Flux units are arbitrary. See Table \ref{table:basic} for the corresponding lines (\hbeta\ or \MgII) and basic properties of each object. The full content of the table is available in the online version. The errors on the LCs are the original PrepSpec output, while we have used inflated 3\% fractional errors in the LCs in our CCF analysis (see \S\ref{sec:lag_det} for details).
\end{tablenotes}
\end{table}

\section{Results}\label{sec:disc}



\subsection{Lags, BH masses, and the $R-L$ relation}

Figs.\ \ref{fig:ccf}--\ref{fig:ccf4} presents the 15 lags in this work, where we show the LCs, the CCF, and the model RMS broad-line profile for each detection. The auto-correlation function (ACF) of the continuum LCs is also shown for comparison with the CCF. {We also show the CCCDs of these lag measurements in Fig.\ \ref{fig:cccd}. In all cases, there is a reasonably well-defined primary peak in the CCCD for us to determine the best lag and associated uncertainties. In a few cases there are sub-structures that are possible aliases due to the sparse sampling of the LCs. These sub-structures in the CCCD tend to increase the uncertainties in the lag measurements.}

We consider these detections as preliminary, with the expectation that adding photometric LCs in our future work will significantly improve these detections (or falsify a small fraction of them, if any). We measure the broad-line dispersion $\sigma_{\rm rms}$ (the second moment) and its uncertainty from the RMS spectra $B_\ell(\lambda)$ (continuum and narrow-line flux modeled and subtracted in individual epochs) produced by PrepSpec. The mean continuum luminosity at rest-frame 5100\,\AA, $L_{5100}\equiv \lambda L_\lambda |_{5100\,\textrm{\AA}}$, is measured from a multifunctional fit to the final coadded spectra of all 32 epochs \citep{Shen_etal_2015a} that models the continuum, broad \FeII\ emission and broad+narrow line emission \citep[e.g.,][]{Shen_etal_2008}. We have estimated the fraction of host starlight contributing to $L_{5100}$ using a spectral decomposition approach described in \citet{Shen_etal_2015b}, and derived the AGN-only continuum luminosity whenever the decomposition is physically meaningful (i.e., the host fraction is non-negative at 5100\,\AA). This last step is important in deriving an unbiased BLR size-luminosity ($R-L$) relation for low-redshift AGN with significant host contamination \citep[e.g.,][]{Bentz_etal_2013}.

We compute a virial product (VP) using the measured time lag and RMS line width: 
\begin{equation}\label{eqn:vp}
{\rm VP}=\frac{c\tau_{\rm cent}\sigma^2_{\rm rms}}{G},
\end{equation}
where $c$ is the speed of light and $G$ is the gravitational constant. The RM BH mass estimate is related to the virial product as $M_{\rm RM}=f{\rm VP}$, where $f$ is the average geometric factor (the virial coefficient) to account for the difference between line width and the virial velocity \citep[e.g.,][]{Shen_2013,Peterson_2013}. The value of $f$ is usually determined empirically using the local BH mass -- bulge stellar velocity dispersion ($\sigma_*$) relation and measurements of $\sigma_*$ in a subset of RM AGN \citep[e.g.,][]{Onken_etal_2004,Graham_etal_2011,Park_etal_2012,Grier_etal_2013,Ho_Kim_2014}. We adopt a fiducial value of $f=5.5$ for this work \citep{Onken_etal_2004}, which is consistent with the latest work \citep{Ho_Kim_2014}. We emphasize that this is the average $f$ factor, and does not capture the diversity in the orientation and BLR structure in AGN \citep[e.g.,][]{Shen_Ho_2014,Pancoast_etal_2014}. The scatter in the actual virial factor for individual objects is currently a large contributor (a factor of 2-3) to the overall systematic uncertainty in RM BH masses \citep[e.g.,][]{Peterson_2013}.

We can also compute a single-epoch (SE) BH mass estimate using the quasar continuum luminosity and the broad-line \hbeta\ FWHM measured from the mean spectrum:
\begin{eqnarray}
&&\log \left(\frac{M_{\rm SE,H\beta}}{M_\odot}\right)= \nonumber\\
&&a+b\log \left(\frac{L_{\rm 51000,tot}}{10^{44}\,{\rm erg\,s^{-1}}}\right) + 2\log{\rm \left(\frac{FWHM_{mean,H\beta}}{km\,s^{-1}}\right)}\ ,
\end{eqnarray} 
with fiducial coefficients of $a=0.91$ and $b=0.5$ \citep[][]{Vestergaard_Peterson_2006}, calibrated against the RM BH masses with $f=5.5$ in the low-$z$ RM AGN sample. The total luminosity at restframe 5100\,\AA\ is used here instead of the AGN-only luminosity because the former was used in the original calibration. While the dynamic range in BH mass is narrow in our sample, we found that the SE masses based on \hbeta\ are grossly correlated with RM masses with a scatter of $\sim 0.28$ dex, consistent with the results in \citet{Vestergaard_Peterson_2006} based on the local RM sample. However, there is a systematic offset of $\sim 0.16$ dex between the two sets of masses, indicating that there may be systematics when applying the \citet{Vestergaard_Peterson_2006} SE mass recipe to our objects at higher redshifts, or in the details of measuring the RMS line widths. This minor discrepancy deserves further investigation with more RM results. 

While not directly used for the calculations in this work, we also include in Table \ref{table:basic} measurements of the line dispersion measured from the mean spectrum and the FWHM measured from the RMS spectrum for each line with a lag for completeness. These were measured from the PrepSpec models of the broad components of the \hbeta\ and \MgII\ emission lines. Two concerns with measuring the line dispersion are (a) the possibility of a bias introduced from blending in the wings with other emission components \citep[see][]{Denney_etal_2009a}, and (b) where to set the line boundaries. These are not a concern here because we are measuring the widths from the model profiles. This makes internally consistent measurements within our analysis, but they are model dependent and therefore may be difficult to reproduce if different assumptions are made for the spectral decomposition. BH mass estimates using alternative SE estimators can be obtained with the spectral measurements reported in Table \ref{table:basic}.


Fig.\ \ref{fig:rl_relation} shows the BLR size-luminosity relation (the $R-L$ relation) based on the rest-frame 5100\,\AA\ quasar-only luminosity, where we compare our lag detections to the compilation of low-$z$ RM results from \citet{Bentz_etal_2013}. All low-$z$ RM results are based on the \hbeta\ line. Our detections are at $z\gtrsim 0.3$, but their locations on the $R-L$ plot are consistent with those occupied by the low-$z$ RM AGN sample within the uncertainties. The 3 most luminous objects with \hbeta\ lags seem to fall below the local $R-L$ relation (albeit with large error bars on the lags), which may be an indication that correlated errors in the continuum and line LCs from spectroscopy-alone biased the lag measurements (see \S\ref{sec:lag_det}). Although the selection of the reported detections is not complete (see \S\ref{sec:lag_det}), we did not intentionally choose detections that are consistent with the known $R-L$ relation, and hence there is no obvious bias from our selection of these detections. In addition, the several \MgII\ lags seem to follow the same $R-L$ relation based on \hbeta. The latter observation suggests that, at least in some quasars, there is overlap between the regions in which broad \MgII\ and broad \hbeta\ originate \citep[see also][]{Sun_etal_2015}, which is expected as both lines have similar ionization potentials. Of course, more data and analysis are needed to test this scenario. 

The apparent flatness in the $R-L$ relation for our detections in Fig.\ \ref{fig:rl_relation} is mostly due to selection effects, and secondly due to the facts that the statistics and the dynamic range in luminosity for our sample are limited and that the lag measurement errors are substantial, rendering a potential correlation statistically insignificant. Given the sampling and duration of our spectroscopic monitoring, we are most sensitive to lags on the order of tens of days \citep{Shen_etal_2015a}. Shorter lags are difficult to detect given the sparse sampling of the spectroscopic LCs, while longer lags are difficult to detect given the limited temporal baseline. A more detailed analysis of the $R-L$ relation based on SDSS-RM lag detections is beyond the scope of this work, which will require more lag detections and proper treatment of selection biases induced by our program, and will be the focus of future SDSS-RM publications. For example, short lags may be recovered when incorporating more densely sampled photometric light curves, or at least upper limits can be placed and used in quantifying the $R-L$ relation. On the other hand, simulations can be used to quantify the completeness in lag detections as a function of lag, given the parameters of the SDSS-RM program \citep{Shen_etal_2015a}. 

\begin{figure}
\centering
    \includegraphics[width=0.48\textwidth]{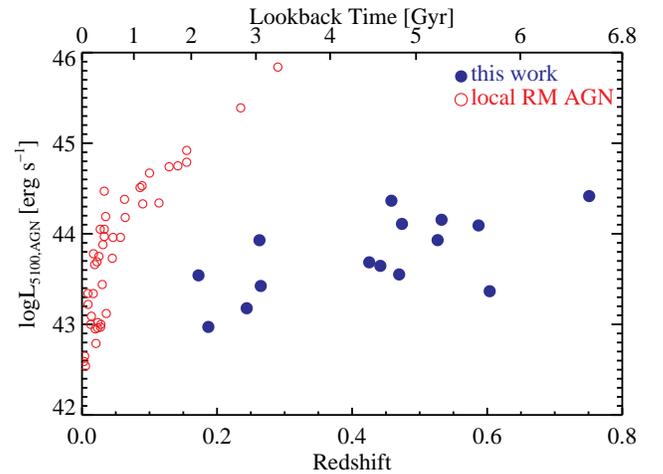}
     \caption{Distribution of objects with detected lags in the redshift-luminosity plane. The red open circles are the 44 local RM AGN compiled in \citet{Feng_etal_2014}, and the blue filled circles represent the 15 preliminary lag measurements in this work. Our lag detections probe a new regime in this parameter space, providing direct SMBH masses over $\sim$ half of cosmic time. }
    \label{fig:Lz_dist}
\end{figure}

Finally, Fig.\ \ref{fig:Lz_dist} demonstrates the improvement of our results in the redshift-luminosity coverage of RM experiments. Our lag detections probe a new regime in this parameter space, providing RM measurements over $\sim$ half of cosmic time. {The median redshift is 0.03 for the local RM sample and 0.46 for our sample. }

\subsection{Additional Notes on Individual Objects}\label{sec:notes}

While our analysis demonstrates that the majority of these lag measurements are true detections, prior reverberation studies \citep[e.g.,][]{Peterson_etal_2004} indicate that when working with data that have low sampling rates, our lag detection methods can sometimes yield incorrect lag measurements despite reporting formally small uncertainties. The criteria for a robust measurement are difficult to quantify, and thus such classifications are somewhat subjective in nature, as they often rely on being able to ``see'' the reverberation signatures and having a narrow, well-defined peak in the CCF and/or CCCD. As such, our prior experience with reverberation mapping leads us to identify a few of our lag measurements as lower-confidence measurements due to various indicators, such as the lack of visible reverberation signatures in the light curves and/or a poorly-defined or low peak correlation coefficient CCF. While our false-positive tests indicate that the detected time delays are likely real (and the lag measurements themselves are also supported using the JAVELIN method), we suspect that the accuracy of a few of the detections may be compromised for various reasons, despite the fact that they passed all of our statistical and quantitative tests for lag detections. 

The two targets where it is most difficult to evaluate the accuracy are RMID 229 and 589. There are very few features visible in the continuum light curve for RMID 229; the two most striking features are due to single epochs that are low S/N and were removed from several other targets due to a suspect calibration, though they were not identified as outliers for this particular target. If the line light curve is shifted by the best-determined-lag from our cross-correlation analysis to see if the continuum and line light curve features line up after this shift, the match is not extremely apparent, and the measured lag places many variability features observed in the line into gaps between observations of the continuum. In addition, the peak of the CCF is low ($\sim 0.6$).  The CCF for RMID 589 is extremely broad and flat-topped; most of the correlation is driven by a long-term increase in flux over the entire set of observations rather than by shorter-term variability features that are more indicative of BLR reverberation. There is again a poor match with the continuum variability when the line light curve is shifted by the measured lag.  

For both RMID 191 and RMID 320, the correlation seems to be driven by either or both of the two possibly-suspect epochs (epochs 3 and 7) and excluding those two epochs (at the cost of losing temporal sampling further) makes the correlation noticeably worse (albeit still formally consistent) in both of these targets. Again, when the line light curves are shifted by the measured time lag, the match is mediocre; the variability signal in the continuum of RMID 191 is rather low amplitude, and it is difficult to see similar features shared by both the line and continuum light curves. The peak of the RMID 191 CCF is also $\lesssim 0.6$, which is lower than that seen for our higher-confidence targets.   

RMID 694 and RMID101 are cases that are somewhat marginal, but should be improved with the photometry and/or additional years of data. RMID 694 has inopportune gaps in the data; there is a gap in the spectroscopy in the middle of one of the major features of the light curve that is potentially biasing the cross-correlation analysis --- the incorporation of the photometry should resolve this issue. RMID 101 has a visible feature in the continuum that we only begin to see reverberated in the line when our campaign ended -- thus, the lag measurement is likely more uncertain than it would be had we continued to observe for another month. 

We anticipate that the ambiguities and concerns listed above will be resolved when the photometry and/or additional monitoring beyond the first six months is included in the analysis for these targets. Results for the remaining nine lag detections appear more robust; the light curves show a better match when the line light curve is shifted by its measured time delay. In addition, they generally have higher peak correlation coefficients in their CCFs (above 0.6), narrower and less flat-topped CCFs, and in most cases, the eye can pick out matching features in the continuum and line light curves that indicate a more robust, accurate lag measurement. 

\section{Discussion and Conclusions}\label{sec:con}

From its theoretical definition by \citet{Blandford_McKee_1982}, it took many years for
reverberation mapping to reach its current state with some 60 
local systems with well-measured lags, sometimes for multiple emission
lines \citep[e.g.,][]{Peterson_etal_2004}. RM studies of local AGN are increasingly focused
on detailed studies of individual objects to use the velocity
dependence of the lags to study the structure of individual
BLRs in detail \citep[e.g.,][]{Bentz_etal_2010b,Grier_etal_2013b}.  It is not possible, however, to increase the
overall size of the sample by an order of magnitude using the
object-by-object approach that has been so successful to date.

The only way to increase greatly the sample size and to start
to probe the UV emission lines of higher redshift systems is 
to use the multiplex advantage of multi-object spectrographs.
This has been explored theoretically in the contexts of both
the SDSS survey using the SDSS-BOSS spectrograph \citep{Shen_etal_2015a} and the
Dark Energy Survey (DES) using the AAOmega spectrographs \citep{King_etal_2015}.  These two studies
emphasized slightly different goals but outlined the survey
durations and cadences likely to yield transformative ($\gtrsim 100$)
numbers of new lag measurements across a broader range of
luminosities, redshifts, and emission lines (e.g., \hbeta, \MgII, and \CIV).   

In both cases, these theoretical studies led to observational
programs, and here we report initial results from the first year of
spectroscopic observations in the SDSS-RM program. Compared to the predictions outlined in the theoretical studies, the
allocated SDSS-RM spectroscopic observations {\it alone} for this first
exploratory study appear sub-critical, given the complexities in real
light-curve data that were difficult to model in the simulations. On the other hand, the SDSS-RM
program has obtained a large amount of supplementary photometric
monitoring data from CFHT Megacam and the Steward Observatory 
Bok telescope that will be used to provide better sampled 
continuum light curves to compare to the line light curves
from the spectroscopic data in our upcoming analyses.
It is important, however, to show the results from simply
analyzing the spectroscopic data alone for two reasons. First,
even with this incomplete data set, we can measure a reasonable number of lags, thereby directly
demonstrating the promise of the approach. Second, the yields based on spectroscopy alone
are low and the uncertainties are great, as predicted
by the theoretical studies for the as-obtained duration and 
cadence. It is desirable to increase the intensity of monitoring in order to have a significant yield of high precision lag measurements in future multi-object RM programs.

In this study, we have presented preliminary lag measurements for 15 quasars at $z<0.8$ included in the SDSS-RM project, using the 6 months of 2014 spectroscopy alone. Our targets are fainter by 1-2 orders of magnitude than the local AGN that have been monitored for RM purposes, highlighting the difficulties of performing RM in this regime. Most of these lags are at $z\gtrsim 0.3$ for intermediate-luminosity quasars, a $L-z$ regime never explored in past work. These results demonstrate the general feasibility of our multi-object RM approach, and provide some confidence for similar programs with multi-object spectroscopy \citep[e.g.,][]{King_etal_2015}. The lags are consistent with 
measurements used for local estimates of the $R$-$L$ relations, but are
not yet available in large enough numbers, precision, or dynamic
range to improve on these local estimates. Several of them are,
however, for the \MgII\ line, and demonstrate that \MgII\ does reverberate
similar to \hbeta. A recent line variability study using SDSS-RM data also found that \MgII\ does vary with similar (albeit slightly smaller) amplitudes to \hbeta\ \citep{Sun_etal_2015}, as supported by the \MgII\ lag detections presented here. Moreover, we can see statistically that we are
close to measuring a much larger number of lags (see Fig.~\ref{fig:diag}).  The
problem is that with the present cadence and duration, there are many ambiguities in how to overlap the line and continuum
light curves to produce an unambiguous lag estimate unless the 
structure of the variations is optimal, and this occurs for only a
small fraction of the targets with {\em spectroscopic-only} light curves.  

For the present SDSS-RM program we will address this challenge in 
future papers by adding the better sampled continuum light curve data
from our imaging programs (e.g., more than doubling the data points from spectroscopy). This should greatly increase the light
curve yield. Denser photometric light curves will enhance the cross-correlation signal, and provide a remedy for correlated errors between the continuum and line flux measurements from spectroscopy.\footnote{While correlated errors in the spectroscopic-only LCs do not seem to affect our reported lag detections much (as confirmed by the discrete-correlation-function analysis), we did observe some cases where there is a strong peak in the CCF near zero lag, and a second peak close to the expected lag from the $R-L$ relation (see \S\ref{sec:lag_det} and Fig.\ \ref{fig:diag} for details). The addition of photometric LCs will help remove these spurious zero-lag peaks in the CCF and recover the true lag. } Our preliminary investigation on the photometric data showed great promises on this improvement. 

{It is also important to remind the reader that even for the small fraction of objects and the spectroscopic-only data we analyzed here, the reported lags are an initial demonstration rather than a final conclusion on the yield of spectroscopic-only lags. There are many more formally detected lags based on the quantitative criteria in \S\ref{sec:lag_det} that require refinements with photometric data. Moreover, we already see multiple-line lag detections in the same objects with this partial data set, as well as lags for other broad lines (such as \halpha, \HeII, etc.), but will defer such focused studies to upcoming SDSS-RM publications. }


Extending the SDSS-RM program, particularly by adding more seasons of spectroscopic data would also greatly improve
the overall impact of the program. The SDSS-RM project continues to perform photometric and spectroscopic monitoring (albeit at reduced cadences) of the target field, which will strengthen the preliminary lag detections and reduce their uncertainties with more data. We have already observed another 12 spectroscopic epochs with the SDSS BOSS spectrograph in the 2015A semester (over 6 months) within the extended Baryon Oscillation Spectroscopic Survey \citep{Dawson_etal_2015}, with 2 epochs each month using a nominal exposure of 1 hr per epoch. Photometric monitoring in 2015 was carried out with a weekly cadence over the same period on the CFHT and Bok telescopes. Additional monitoring will be sought in 2016 and beyond. With the extended multi-year spectroscopic time baseline and earlier photometric light curves from PanSTARRS in the SDSS-RM field \citep{Kaiser_etal_2010,Shen_etal_2015a}, we will be able to expand RM lag detections to the high-luminosity and long-lag regime at $z>1$, which will further test the practical value of RM at high $z$ and better constrain the $R-L$ relations as functions of redshift, line species, and quasar properties.

\acknowledgements

{We thank the referee for a careful review and constructive comments.} Support for the work of YS was provided by NASA through Hubble Fellowship grant number HST-HF-51314, awarded by the Space Telescope Science Institute, which is operated by the Association of Universities for Research in Astronomy, Inc., for NASA, under contract NAS 5-26555. KH acknowledges support from UK Science and Technology Facilities Council (STFC) grant ST/M001296/1. CJG and WNB acknowledge support from NSF grant AST-1517113 and the V.M. Willaman Endowment. BMP is grateful for support from the National Science Foundation through grant AST-1008882. KDD is supported by an NSF AAPF fellowship awarded under NSF grant AST-1302093. JRT acknowledges support from NASA through Hubble Fellowship grant HST-HF-51330 awarded by the Space Telescope Science Institute, which is operated by the Association of Universities for Research in Astronomy, Inc., for NASA under contract NAS 5-26555. MS acknowledges support from the China Scholarship Council (No. [2013]3009).

Funding for SDSS-III has been provided by the Alfred P. Sloan Foundation, the
Participating Institutions, the National Science Foundation, and the U.S.
Department of Energy Office of Science. The SDSS-III web site is
http://www.sdss3.org/.

SDSS-III is managed by the Astrophysical Research Consortium for the
Participating Institutions of the SDSS-III Collaboration including the
University of Arizona, the Brazilian Participation Group, Brookhaven National
Laboratory, University of Cambridge, Carnegie Mellon University, University
of Florida, the French Participation Group, the German Participation Group,
Harvard University, the Instituto de Astrofisica de Canarias, the Michigan
State/Notre Dame/JINA Participation Group, Johns Hopkins University, Lawrence
Berkeley National Laboratory, Max Planck Institute for Astrophysics, Max
Planck Institute for Extraterrestrial Physics, New Mexico State University,
New York University, Ohio State University, Pennsylvania State University,
University of Portsmouth, Princeton University, the Spanish Participation
Group, University of Tokyo, University of Utah, Vanderbilt University,
University of Virginia, University of Washington, and Yale University.

\end{document}